\documentclass{iopjournal}
\usepackage{graphicx}
\usepackage{float}
\usepackage{mathabx}
\usepackage{makecell}
\usepackage[square,sort&compress,numbers]{natbib}

\usepackage{amsmath}
\usepackage{orcidlink}
\usepackage{sidecap}
\usepackage{paralist}   
\definecolor{tangerine}{rgb}{1.0, 0.6, 0.4}
\definecolor{forestgreen}{HTML}{228b22}

\begin{document}
\pagestyle{plain} 
\title{Identifying Cost-Favorable Locations for Cosmic Explorer}

\author{Laurence Datrier$^1$\orcid{0000-0002-0290-3129}, Geoffrey Lovelace$^1$\orcid{0000-0002-7084-1070}, Tooba Ansar$^2$, Lance Blagg$^3$\orcid{0009-0009-7388-6459}, Warren Bristol$^4$\orcid{0009-0008-6058-4964}, Matthew Evans$^5$\orcid{0000-0001-8459-4499}, Chris Lukinbeal$^4$\orcid{0000-0003-1827-7764},
Vuk Mandic$^6$,
Kiet Pham$^6$\orcid{0000-0002-7650-1034}, Jocelyn Read$^1$\orcid{0000-0002-3923-1055},
Sarmad Rameez$^7$, Amber Romero$^1$, Oscar Romero$^{1,4}$, Babatunde Isaac Rotimi$^7$\orcid{0000-0003-2184-3077}, Joshua B.~Russell$^7$, Andrew Saenz$^1$, Fran\c{c}ois Schiettekatte$^8$\orcid{0000-0002-2112-9378}, Robert Schofield$^3$, David H. Shoemaker$^5$\orcid{0000-0002-4147-2560}, Bretton Simpson$^1$\orcid{0000-0003-3861-0122}, Bram J.J.~Slagmolen$^9$\orcid{0000-0002-2471-3828}, Joshua R.~Smith$^1$\orcid{0000-0003-0638-9670} }

\affil{$^1$The Nicholas and Lee Begovich Center for Gravitational-Wave Physics and Astronomy, California State University, Fullerton, 92831, USA}\\
\affil{$^2$Department of Physics \& Astronomy, Amherst College, Amherst, Massachusetts 01002, USA}\\
\affil{$^3$Department of Physics, University of Oregon, Eugene, Oregon 97403, USA}\\
\affil{$^4$School of Geography, Development and Environment, the University of Arizona, Tucson, Arizona 85721, USA}\\
\affil{$^5$MIT Kavli Institute, Massachusetts Institute of Technology, Cambridge, Massachusetts 02139, USA}\\
\affil{$^6$School of Physics and Astronomy, University of Minnesota, Minneapolis, Minnesota 55455, USA}\\
\affil{$^7$Department of Earth and Environmental Sciences, Syracuse University, Syracuse, New York 13244, USA}\\
\affil{$^8$D\'epartement de Physique, Universit\'e de Montr\'eal, Montr\'eal, Qu\'ebec, Canada}\\
\affil{$^9$OzGrav, Australian National University, Canberra, Australian Capital Territory 0200, Australia}

\email{ldatrier@fullerton.edu}


\begin{abstract}

Cosmic Explorer (CE) is a proposed next-generation gravitational-wave observatory that aims to extend our gravitational-wave vision to the edge of the observable universe. With a foundation of technology proven by the National Science Foundation's Laser Interferometer Gravitational-Wave Observatory (LIGO), CE will observe black holes and neutron stars across cosmic time, explore the nature of extreme matter with high fidelity, and probe the nature of gravity and fundamental physics.
CE's reference design consists of two widely separated L-shaped detectors to be located in the conterminous United States, one with 20\,km arms and one with 40\,km arms. 
As of 2026, CE is in its design and site evaluation phase, with plans to begin observing in the early 2040s together with the Einstein Telescope in Europe. 
The size of CE observatories---up to an order of magnitude larger than the 4\,km LIGO observatories---presents a significant challenge for identifying suitable candidate sites where CE will achieve its science goals, be built within cost boundaries, attract and retain a workforce, and align with community values. 
In this paper, we report on the design and use of a Python package, the Cosmic Explorer Location Search (CELS) package, to identify cost-favorable sites for CE. 
For a specified detector location and L-shaped geometry in the conterminous United States, CELS estimates site-preparation costs associated with excavation, land clearing, and land acquisition, while accounting for the scientific effects of detector tilt, arm length, and arm opening angle.
After describing the package's methods, we present results for a national-level cost and positioning analysis that complements a recent national suitability analysis. We also discuss how future improvements to CELS will allow deeper, more local studies as the Cosmic Explorer team narrows its list of potential locations. 
\end{abstract}

\section{Introduction}\label{sec:intro}

\begin{figure}
    \centering
    \includegraphics[width=0.6\linewidth]{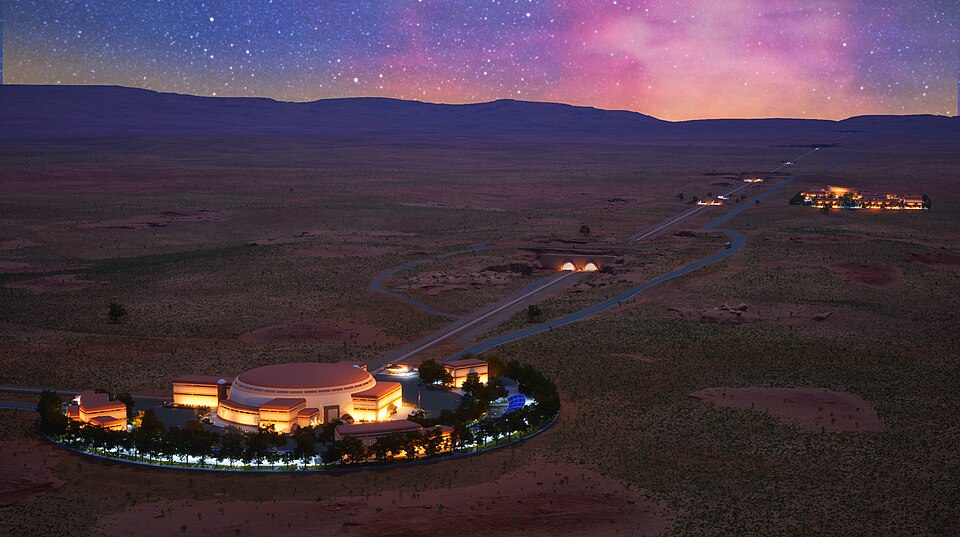}
    \includegraphics[width=0.39\linewidth]{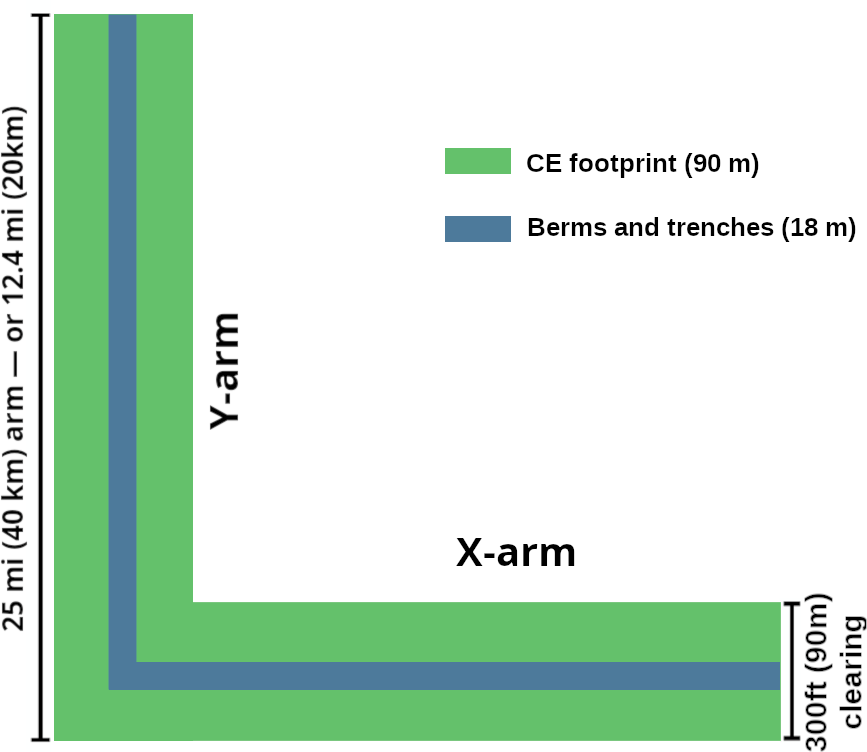}
    \caption{Left: Artist's conception of Cosmic Explorer~\cite{AJenkinsCENight}. The image shows a view of the corner station, looking towards the x-arm. Right: Simplified footprint of a Cosmic Explorer observatory. Note that the simplified clearing shown would include trenches, berms, roads, and observatory infrastructure.}
    \label{fig:CEandFootprint}
\end{figure}
The Laser Interferometer Gravitational-Wave Observatory (LIGO)~\cite{aligo} observatories in Livingston, Louisiana and Hanford, Washington---together with their European partner Virgo~\cite{avirgo}---have inaugurated gravitational-wave astronomy by observing hundreds of signals from merging black holes. They have also enabled the first joint detection of gravitational-wave and electromagnetic signals, with the detection of the well-localized binary neutron star merger GW170817. To date, the LIGO--Virgo--KAGRA collaboration has detected 390 gravitational-wave signals from compact binary coalescences~\cite{LIGOScientific:2016aoc,LIGOScientific:2017vwq,LIGOScientific:2017ync,LIGOScientific:2007fwp, theligoscientificcollaboration2026gwtc50introductionversion50}. 

Cosmic Explorer (CE) is a planned next-generation gravitational-wave observatory
that will extend our gravitational-wave astronomical reach to the edge of the observable universe~\cite{evans2023cosmicexplorersubmissionnsf} by scaling up LIGO's technology and approach.
CE will observe black holes and neutron stars across cosmic time, measure the properties of extreme matter (particularly neutron star matter) with high precision, and will explore the nature of gravity and fundamental physics~\cite{2021arXiv210909882E}.

The CE project is, at the time of writing, in its conceptual/development design and site identification and evaluation phase~\cite{Shoemaker_2026}. 
CE's reference design consists of two widely separated observatories that would be located in the conterminous United States: one with 40-km-long arms (CE40), and one with 20-km-long arms (CE20).
Each CE observatory will have one detector---initially an enhanced (dual-recycled, Fabry-Perot) Michelson interferometer---scaled up from the successful LIGO design. The left panel of figure~\ref{fig:CEandFootprint} shows an artist's impression of a 40\,km CE observatory.
Cosmic Explorer aims to begin observing in the early 2040s, joining a global network of next-generation gravitational-wave observatories that will include the Einstein Telescope~\cite{Punturo:2010zz}---a European underground observatory with two reference configurations, a 10-km-long triangle or two 15-km-long L-shaped detectors, and three candidate sites:  \begin{inparaenum}[(i)] \item the border area of Belgium, Germany and the Netherlands; \item Sardinia, Italy; and \item Lusatia, Germany\end{inparaenum}---and LIGO-India~\cite{Saleem_2022, ligoindiadcc}---a 4\,km LIGO Observatory in Aundha in Maharashtra, India, at either A+ or A\# sensitivity.

Gravitational-wave observatories like CE have different siting requirements from ground-based electromagnetic observatories; in principle, CE could be sited anywhere in the United States, not being bound by siting restrictions related to astronomical seeing. However, the detectors' very long (20\,km and 40\,km) arms, which must be straight and level to accommodate their laser beams' paths, pose unique challenges for identifying potential sites for the CE observatories. In the case where CE40 and CE20 would both be built, the interferometers would also need to be widely separated ($>$1000\,km). Criteria for identifying suitable sites for CE are outlined in~\cite{10.1063/5.0242016}. Suitable sites for Cosmic Explorer will be places where CE can achieve its science goals; be built within cost boundaries; attract, support, and retain its workforce; and where observatory activities can be aligned with community values~\cite{10.1063/5.0242016,Datrier:2025wjs}. 

The U.S. National Science Foundation (NSF) will determine and run the process for CE's site selection.
With the goal of informing the NSF-led site selection process, we and other members of the Cosmic Explorer Project are working to identify and evaluate potential sites that meet the suitability criteria outlined in~\cite{10.1063/5.0242016}.
This effort includes \begin{inparaenum}[(i)] \item remote suitability analysis using Geographic Information Systems (GIS) and publicly available data; \item visits and relationship building~\cite{Joey}; and \item pending permission, on-site physical and sociocultural suitability assessments.\end{inparaenum}

The remote suitability analysis includes several components. First, the National Suitability Analysis~\cite{Bristol_2026} uses a land-and-people approach, searching within acceptable thresholds across 91 interdisciplinary factors, such as seismic noise at a 1-second period based on USArray data~\cite{Anthony:2022} and federally restricted areas, to find locations at which a CE observatory could be physically accommodated and culturally welcomed, achieve its scientific goals, and attract and retain a robust workforce.   
Members of the Cosmic Explorer Project have completed this analysis using the ArcGIS platform, which integrates geospatial data for analysis and management. Second, an adjacent project determined locations that are not viable because they are 
either too close to buildings, roads, railroads, or other features; or have too much environmental noise~\cite{ToobaSURF}.

In this paper, we present a third component of the remote suitability analysis.
Our overall goal is to identify cost-favorable locations---not to cost a specific location or to carefully cost CE construction, operations, or design. The CE Horizon Study~\cite{2021arXiv210909882E} made initial estimates of CE's civil engineering and construction costs; today, architects and engineers are improving these initial estimates as part of the CE Conceptual Design. These ongoing efforts informed the work we present here.

To achieve this overall goal, we developed and used Cosmic Explorer Location Search (CELS), an open-source Python package~\cite{CELS_codebase} that uses publicly available data (such as elevation, land cover, and land value) to estimate the cost of preparing a Cosmic Explorer site for different detector locations and orientations throughout the conterminous United States.
CELS estimates the impact of a potential detector's location and detector geometry on site preparation costs and science output. %
CELS can generate maps of estimated costs for the whole of the conterminous United States, or focus on specific areas of interest.

CELS builds on significant history in siting and GIS/geospatial efforts in astronomy and physics~\cite{Schoeck_2009,Schilizzi2024,fcc2025vol3, ITERJASS2003,Abreu:2023WI,10.1093/mnras/staa201,HASSAN201776,app11188666} and related to terrestrial gravitational-wave observatories~\cite{Kuns_2020,10.1063/5.0242016,Datrier:2025wjs,Bristol_2026,Amann:2020jgo,iacovelli2026closeevaluatingimpactbaseline}.
The main contribution of this work is a reproducible, open-source, national-scale method for comparing CE site preparation costs and scientific outcomes derived from elevation, land cover, land acquisition, major road crossings, and detector geometry.

The remainder of this paper is organized as follows.
Section~\ref{sec:methods} provides a technical description of CELS, including the data sources, methods to estimate costs, and the impact of positioning and site layout on scientific performance. Section~\ref{sec:results} presents results, including a national search for cost-favorable sites for CE40 and CE20 and significant updates (including new data layers and more accurate costing) over the previous national search results presented in~\cite{Datrier:2025wjs}. We briefly conclude and outline our next steps in section~\ref{sec:conclusion}.

\section{Methods}\label{sec:methods}
We developed the Cosmic Explorer Location Search (CELS) Python package to assist with the identification of cost-favorable sites for Cosmic Explorer.
CELS estimates relative site-preparation costs, not full observatory construction or operating costs.
The goal of CELS is to identify locations where the topography, geography, and geology are favorable to building CE, in the sense of having low estimated site preparation costs. 
CELS produces low-accuracy, early-phase, rough-order-of-magnitude cost estimates (-25\%--+75\%) to support preliminary go/no-go decisions.
In the future, we anticipate that these cost estimates will be refined as part of a feasibility study carried out professionally by architects and engineers.

Earthmoving is a major contribution to the cost of preparing a site for construction of a Cosmic Explorer observatory. Because the laser beams in the detector's arms travel in essentially straight lines, CE's beam tubes must be straight and level. If Earth were perfectly spherical, for instance, in order to build CE along the local geodetic horizon plane, a beam tube of length $L_{\rm arm} = 40~{\rm km}$ whose midpoint rests on Earth would have end stations $L_{\rm arm}^2/8 R_{\Earth} \sim 30~{\rm m}$ above the surface. 
In such a situation, even if an optimal vertical positioning is found for the beam tubes, the cost of earthmoving for digging trenches or tunnels and building berms or bridges to support the straight beam tubes would exceed reasonable project cost estimates. 

From an earthworks perspective, an ideal CE site would be a Euclidean (geometrically) flat plane in three-dimensional space. Because Earth is curved, such a surface corresponds to a shallow bowl shape in elevation coordinates since the plane's center is closer to the center of Earth than its edges. Ideally, this plane would also have little or no overall tilt (i.e., the plane should have zero average slope).
Other significant costs to prepare a site for CE include the costs to acquire and clear the land and costs to cross roads, streams, and wildlife paths (e.g., with bridges or underpasses).

To identify cost-favorable sites, CELS explores different locations for the L-shaped detector corner station (or vertex) as well as different arm lengths, orientations, and opening angles. CELS is an open source project hosted on GitLab that uses publicly available map and GIS data. 
It was originally based on the work outlined in~\cite{Kuns_2020}. Besides adding new features to that previous work (such as estimating the cost of land acquisition), CELS 
also enables data sharing through reading and writing common file types, such as GeoTIFF and ESRI\footnote{The current instance of CELS only takes in these file formats; however, the code can easily be rewritten to take in other, open-source formats as input (and output).} shape files. 
The CELS package also introduced continuous integration (automatic testing), code review, and documentation to improve code quality and facilitate contributions from multiple developers.

In this section, we first summarize the open data sources that we used as inputs to CELS in section~\ref{ss:datasources}, and then describe how CELS estimates site preparation costs in section~\ref{ss:costs}.

\subsection{Data sources}~\label{ss:datasources}
Figure~\ref{fig:data} shows the data layers currently used in CELS. These are drawn from publicly available GIS datasets and include raster layers---bitmaps that assign numerical values to pixels---and vector layers---which use features such as lines---to represent real-world features.  

Spatial geographic data is typically presented in coordinate systems known as map projections that allow the data from the curved surface of the Earth to be displayed on a flat surface. Different map projections have different uses, regions of validity, and associated uncertainty. Spatial data can be transformed from one projection into another using reprojection.  
In CELS, we use EPSG:5070, an Albers Conserved Area map projection, because it is designed for data analysis and small-scale data presentation for the conterminous lower 48 states, and, like much of our analysis, uses meter-based length units~\cite{EPSG:5070}. EPSG:5070 prioritizes the conservation of area over the conservation of angles, leading to some distance distortions ($\mathcal{O}(1\%)$), dependent on orientation and location.  
To prepare the data layers shown in figure~\ref{fig:data} for use in CELS, we reprojected each of them into the EPSG:5070 projection using \textsc{gdal} and the average resampling method, which resamples arc-seconds into average meters per pixel. The data layers used in CELS and their sources are as follows:

\begin{figure}
\centering
\includegraphics[width=1\textwidth]{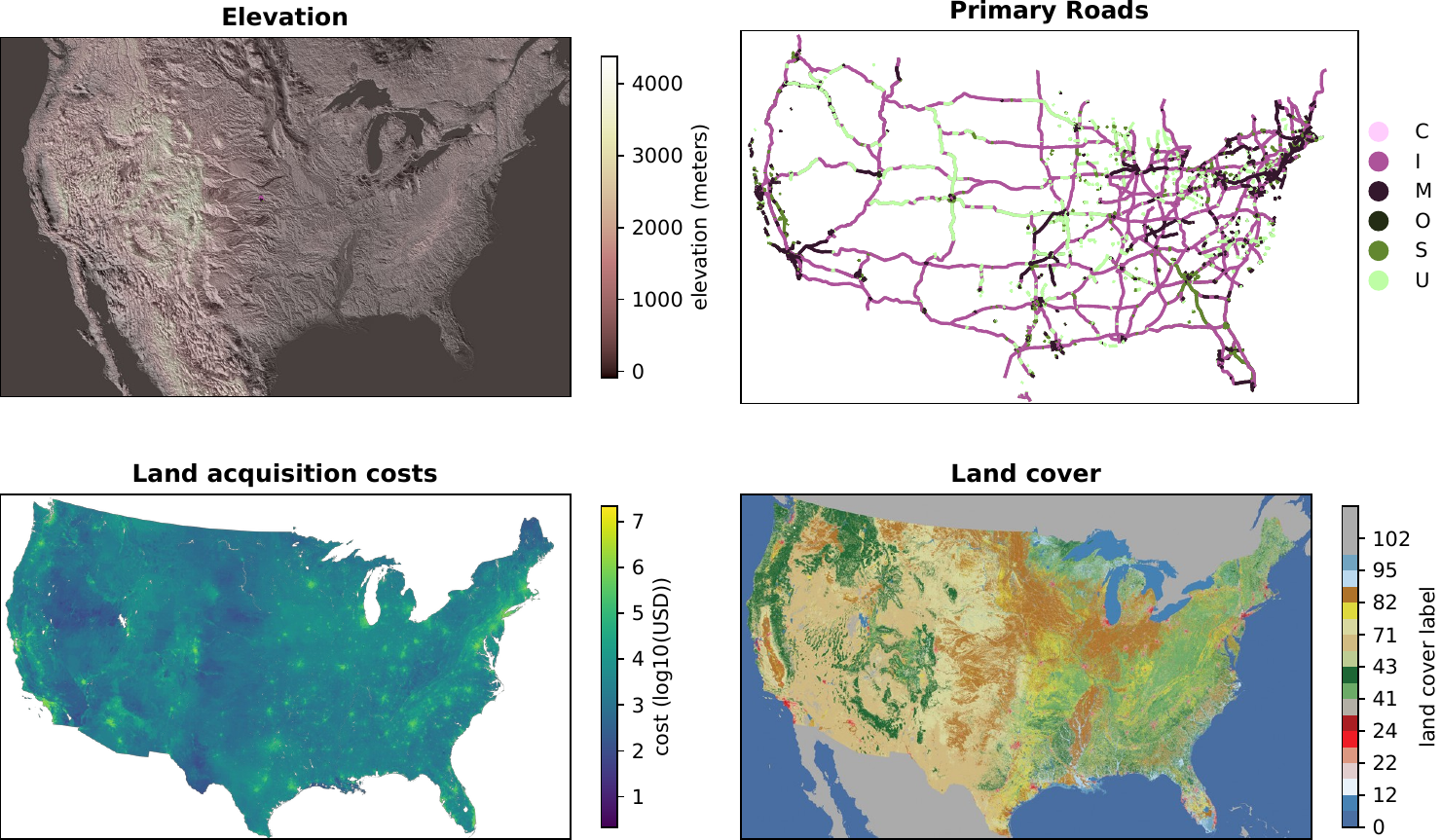}
    \caption{The four data layers used in the CELS national search, projected in EPSG:5070. Land cover labels follow NLCD classifications~\cite{NLCD2021} (details in Table~\ref{table:lc_costs}), and major roads are derived from the MAF/TIGER database~\cite{MTDB} (details in Table~\ref{table:roads}). Elevation data is from the USGS National Map~\cite{NationalMap} in units of meters. Land acquisition costs (USD per acre, log scale) are derived and modified from~\cite{doi:10.1073/pnas.2012865117}. All layers are based on publicly available data as described in the text.}
    \label{fig:data} 
\end{figure}

\paragraph{Elevation} For elevation data, we use the (raster) elevation layer of the USGS National Map~\cite{NationalMap}. This data is a tiled collection of the 3D Elevation Program (3DEP)'s 1 arc-second resolution data presented as a gridded raster with each point's value representing the elevation above the North American Vertical Datum of 1988 (NAVD88).

\paragraph{Land cover} Land cover data spatially maps and describes the physical material on the surface of the land such as forests, open water, or urban development. For land cover, we use the (raster) 30\,m resolution National Land Cover Database (NLCD) 2021 dataset from the U.S.~Geological Survey (USGS)~\cite{NLCD2021}. This dataset assigns a value to each pixel that corresponds to a type (listed in Table~\ref{table:lc_costs}) of land cover.

\paragraph{Land acquisition} We derived the land acquisition cost dataset from the high-resolution (raster) estimated land values presented in \cite{doi:10.1073/pnas.2012865117}, which is based on econometric modeling of approximately six million property sales across the conterminous United States. The source dataset reports values as ln(U.S. Dollars [USD] (2017)) per hectare; 
for this analysis, we converted these values to USD (2017) per acre, while retaining the original 480-meter resolution. To ensure spatial consistency across datasets, the raster extent was expanded to match that of the NLCD layer~\cite{NLCD2021}. We assigned an acquisition cost of 0 USD (2017) per acre to inland water bodies, which were not included in the original data set's land values. 

In post-processing our search results, we accounted for inflation by converting USD (2017) to USD (2026) at a rate of 1\,USD (2017) = 1.37\,USD (2026). This rate is obtained from~\cite{BLSInflationCalculator} and is a simplification of the real property inflation rate. It is important to note that in spatial econometrics, real property values, buying power, farmland and rangeland revenue, and other factors should be taken into account to properly account for inflation. This will be done in future, more localized analyses.

\paragraph{Roads} 
Roads are, in principle, included within the NLCD national land cover dataset---mostly as ``developed, high intensity" areas (see table~\ref{table:lc_costs} for details)---but that dataset does not capture all roads because of its finite resolution (30\,m), and it does not contain detailed road type information. 
Therefore, we use a separate (vector) data layer of roads from the U.S. Census Bureau: primary roads data from the Master Address File/Topologically Integrated Geographic Encoding and Referencing (MAF/TIGER) Database (MTDB)~\cite{MTDB}. We plan to extend our analysis to include railroads in future work.

\subsection{Site-related costs}\label{ss:costs}

\begin{table}[ht]
\centering
\begin{tabular}{lr}
\hline
\hline
\textbf{Parameter} & \textbf{Value}\\  \hline
\hline
Arm length (CE40) &40\,km\\
Arm length (CE20) & 20\,km\\
Footprint width & 300\,ft/91.44\,m \\
Footprint area (CE40) & ${~\sim}$730\,ha/1,800\,ac/7,300,000\,m$^2$\\
Footprint area (CE20) & ${~\sim}$365\,ha/900\,ac/3,600,000\,m$^2$\\
Width of berms and trenches & 18\,m\\
Width of overpasses and underpasses & 16.46\,m\\
Angle of repose & 45\,$^{\degree}$\\
Minimum tunnel depth & 15.1\,m\\

\hline
\hline
\end{tabular}
\caption{Length and area parameters used by CELS.}
\label{tab:CE_parameters}
\end{table}

Along their arms, both Cosmic Explorer observatories will require a cleared width of at least 300\,ft, or ${\sim}$90\,m, to accommodate the beam tube and access roads, giving a total footprint area of ${\sim}$1800 acres for a CE40, or ${\sim}$900 acres for a CE20, as shown in figure~\ref{fig:CEandFootprint}. Table~\ref{tab:CE_parameters} contains details of the length and area parameters used in CELS. Given this footprint, many factors will affect the costs of preparing the CE sites for construction, including the type of land overlapping the observatory's footprint, the monetary land value, obstructions such as roads and railways, and the elevation along the observatory's arms. 
Our goal is to provide rough-order-of-magnitude estimates of site-preparation costs to help identify cost-favorable locations in the U.S. To achieve this, CELS currently includes only the expected leading-order costs, as described below. 
All costs presented below are in millions of U.S. dollars (M USD) for the year 2026---when necessary, adjusted using the inflation calculator provided by the U.S. Bureau of Labor and Statistics~\cite{BLSInflationCalculator}. 

\subsubsection{Land cover}
Preparing the site footprint will require clearing of vegetation, structures, or other types of land cover. 
Table~\ref{table:lc_costs} presents rough (one significant figure) cost estimates, in USD per acre, for clearing land with that type of land cover to construction-ready conditions based on publicly available land clearing project costs. 
Land with little vegetation is associated with lower clearing costs, forests with higher costs, and wetlands---which often require more labor to clear and may have regulations on the type of equipment used---are higher still.
The costs for clearing developed land are based on estimates for demolition costs and clearing. 
For the special case of open water or ocean, the cost given is the estimate for building a bridge. This is prohibitively expensive and effectively excludes sites crossing large bodies of open water. 
Note that we do not currently take into account regional cost factors and economies of scale (clearing 100 acres is less than 100 times as expensive as clearing one acre). 

\begin{table}
\vspace{0.5cm}
\begin{center}
\begin{tabular}{|l|l|r|}
\hline
Label & Land type & Cost (USD) per acre\\
\Xhline{1.2pt}
0&ocean&\$7,000,000 \\
\hline
11 &open water& \$7,000,000\\
\hline
21 & developed, open space &\$6,000\\
\hline
22 & developed, low intensity & \$20,000\\
\hline
23 & developed, medium intensity&\$50,000\\
\hline 
24 & developed, high intensity & \$200,000\\
\hline
31 & barren land (rock/sand/clay)&\$1,000 \\
\hline
41 & deciduous forest& \$6,000 \\
\hline
42 & evergreen forest & \$6,000\\
\hline
43 & mixed forest & \$6,000\\
\hline
51 & dwarf scrub& \$2,000\\
\hline
52 & shrub/scrub&\$3,000 \\
\hline
71 & grassland/herbaceous&\$1,000 \\
\hline
72 & sedge/herbaceous& \$1,000\\
\hline
73 & lichens& \$1,000\\
\hline
74 & moss& \$1,000\\
\hline
81 & pasture/hay & \$1,000\\
\hline
82 & cultivated crops&\$1,000 \\
\hline
90 & woody wetlands & \$8,000\\
\hline
95 & emergent herbaceous wetlands&\$8,000 \\
\hline
\end{tabular}
\end{center}
\caption{Land cover types and costs in USD. Not included in this table are labels for areas outside the conterminous United States (i.e., Mexico, Cuba, and Canada) and land cover types that are not present in the contiguous U.S. (i.e., perennial ice/snow). CELS assigns a prohibitively high cost to these areas in order to essentially exclude them. Cost estimates are derived from contractor pricing estimates.}
\label{table:lc_costs}
\end{table}

\begin{figure}[h]
\centering
\includegraphics[width=1\textwidth]{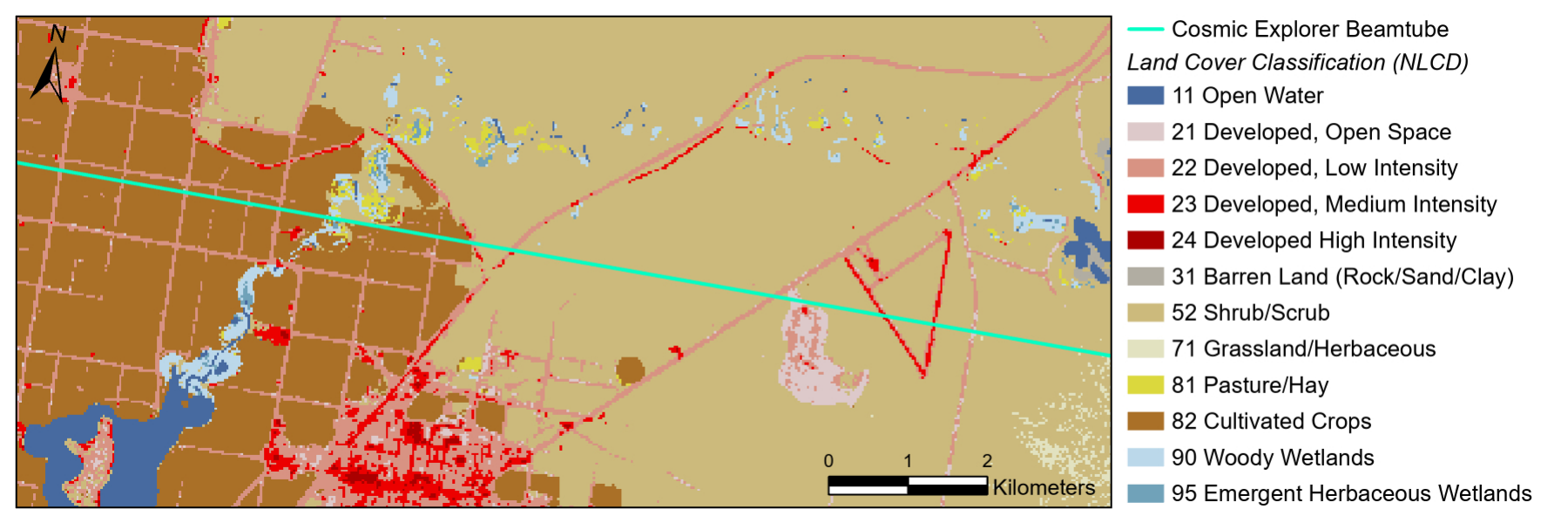}
    \caption{An illustrative example map showing a 15\,km portion of a Cosmic Explorer beam tube (aqua line) overlapping different types of land cover.}
    \label{fig:lc} 
\end{figure}

To estimate the costs associated with clearing and preparing a site with a given type of land cover, CELS calculates the number of acres in each length step along the arm, $\Delta l$, by multiplying by the planned width of Cosmic Explorer's site footprint $w_{\mathrm{fp}}=91$\,m, and by $A$ to convert m$^2$ into acres, and multiplying by the cost for the corresponding land cover type $C_{\mathrm{type}_k}$ in Table~\ref{table:lc_costs} and dividing by 1 million to give costs in units of 1M USD, 

\begin{equation}
    C_{\mathrm{lc}} = 1\times 10^{-6}  \Delta l \times w_{\mathrm{fp}} \times A \sum_k C_{\mathrm{type}_k}. 
\end{equation}\label{eqn:Clc} 
Figure~\ref{fig:lc} shows an illustrative example of a portion of a detector arm and the land cover pixel values underneath it.

\subsubsection{Land acquisition costs}

The manner in which land for Cosmic Explorer will be acquired is not yet determined; we expect that the U.S. federal government and the U.S. National Science Foundation would lead land acquisition efforts for CE. Possible models include land purchase, land lease, and temporary grant of land rights and/or right of way. 

For purchase or lease models, the price might vary regionally with land value. 
CELS estimates land acquisition costs by sampling the land value GeoTIFF---which is already in USD per acre---directly for each k-th step $C_{\mathrm{la}_k}$ and summing along the arms, similarly to the land cover calculation, as
\begin{equation}
    C_{\mathrm{la}} = 1\times 10^{-6}\Delta l\times w_{\mathrm{fp}}\times A \sum_k C_{\mathrm{la}_k}.  
\end{equation}\label{eqn:Cla}

\subsubsection{Road crossings}
Roads present a cost and complexity challenge for CE construction. 
The length of CE means that its beam tubes might need to cross existing roads (and railroads). 
Depending on the arm's depth with respect to grade at locations where a CE arm and a road intersect, 
continuing the arm and service roads straight (``as the laser flies'') at such intersections would require an 
overpass or underpass. 
Each road crossing would thus carry additional costs and potential delays and other challenges associated with traffic rerouting and the construction of overpasses or underpasses.  

CELS identifies any road crossings as intersections between a CE detector's arms and the road lines from the vector map layer described in section~\ref{ss:datasources}. For each intersection, a crossing cost is assigned based on the road type, as in Table~\ref{table:roads}. Then we calculate the total road crossing cost $C_{\mathrm{rc}}$ as a sum over all of the individual road crossing costs:
\begin{equation}
    C_{\mathrm{rc}} = \sum_n^{N_c} C_{\mathrm{rc}_n}.  
\end{equation}\label{eqn:Crc} We plan to extend CELS to also include railroad crossing costs, 
but CELS currently only estimates the cost of road crossings.

Table~\ref{table:roads} gives estimated crossing costs by road type that are based on rough order-of-magnitude parametric total project costs, specifically costs for two-lane grade separation overpass or underpass. The class of road being crossed scales the assumed structure length and width for the crossing, with interstates being the longest and county roads being the shortest.
For these estimates, we assume Cosmic Explorer is the ``crossing road" that passes over or under the given type of road; we also assume that the crossing road is 54\,ft in width (${\sim}16$\,m), encompassing two 12-foot travel lanes and shoulders.
For example, an overpass for an interstate assumes a 54\,ft width, 200\,ft length, for an area of 10,800\,ft$^2$. 
We estimate unit costs in $\$/\rm{ft}^2$ for bridging~\cite{CABridgeCosts} and two-lane road elements~\cite{FLDOTcosts}. We assume a rural or suburban setting, which will have limited to moderate costs associated with existing utilities and adjacent constraints, right-of-way, and traffic control. This assumption is justified, since more urbanized settings tend to be excluded by CELS because of their high land cover cost estimates. Road costs given here represent an average of the lowest cost rural numbers and the highest cost suburban numbers. We calibrated these costs against several U.S.~overpass and underpass projects with known construction costs.   

Whether a CE arm is above or below grade at the location of a road crossing determines whether the crossing would require an overpass or underpass. Underpasses are typically more expensive, because they require more excavation and drainage and face more potential utility conflicts. Currently, CELS does not attempt to determine whether a crossing would require an underpass or overpass; instead, CELS just uses the higher cost estimate (i.e., the underpass estimate). We also note that because often, but not continuously, the land cover data classifies roads as areas of dense development, CELS might also count the cost of crossing a road in our land cover cost estimates.

\begin{table}
\vspace{0.5cm}
\begin{center}
\begin{tabular}{|c|l|c|c|c|}
\hline
Label & Road type & Length (ft) & \makecell{Overpass cost \\ (1M USD)} & \makecell{Underpass cost \\ (1M USD)} \\
\Xhline{1.2pt}
C & County & 80 & 9 & 16\\
\hline
I & Interstate& 200& 24 & 58\\
\hline
M & Common Name& 90 & 11 & 19\\
\hline
O & Other& 110 & 12 & 24\\
\hline
S & State& 120 &  14 & 28\\
\hline
U & U.S.& 140 & 16 & 34\\
\hline
\end{tabular}
\end{center}
\caption{Table of road type following the RTTYP (Route Type Code) classification scheme and the rough estimated costs of constructing an overpass or underpass, based on a parametric model described in the text.}
\label{table:roads}
\end{table}

\subsubsection{Elevation}~\label{ss:elevation}

\begin{figure}[h]
    \centering
    \includegraphics[width=1\textwidth]{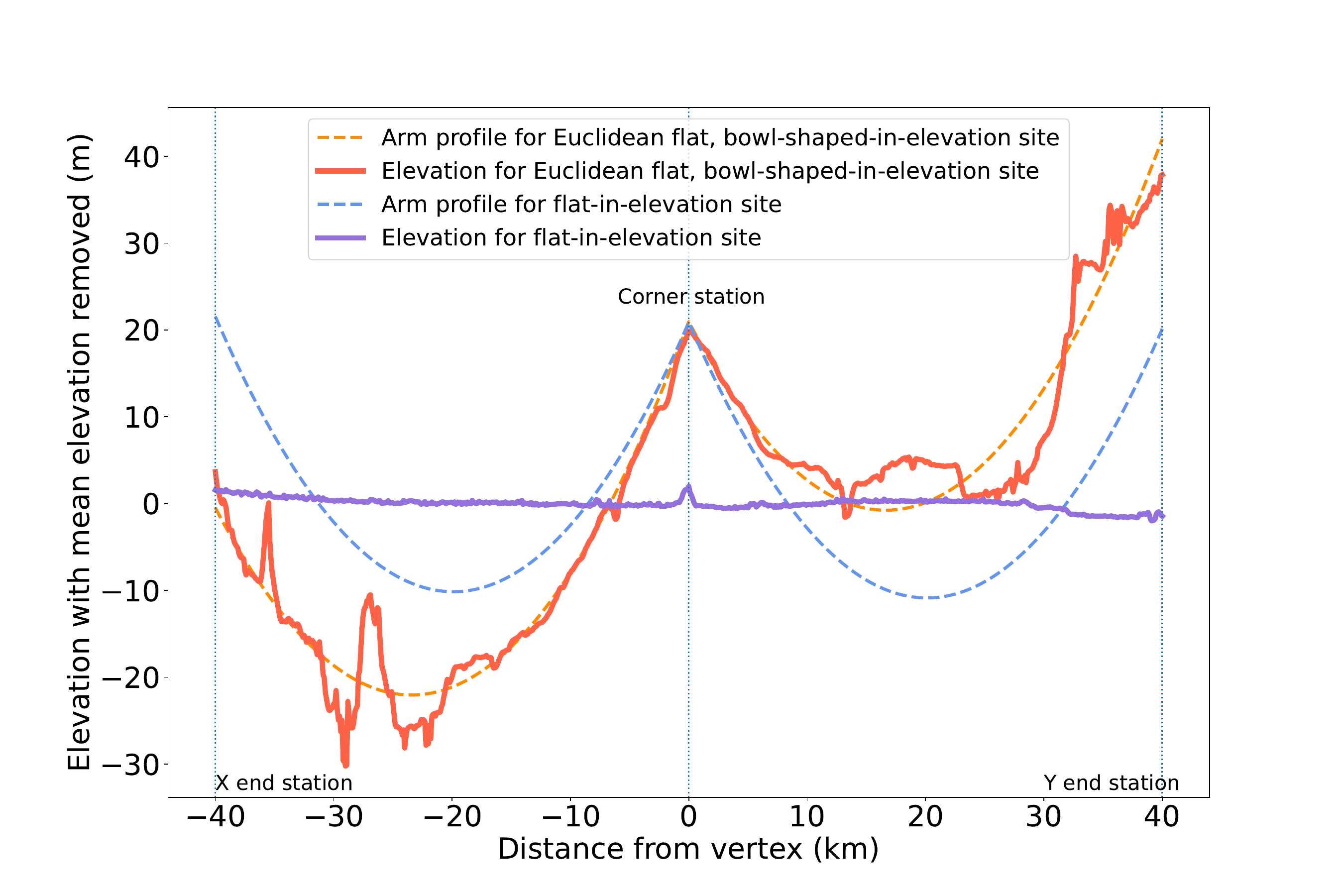}
    \caption{Elevation and arm profiles with mean elevation removed for two locations, one in a constant-in-elevation site, and one in a bowl-shaped in elevation, Euclidean-flat valley. For the Euclidean-flat site, the excavating/build-up volume is ${\sim}$4M\,m$^{3}$, while for the flat site, that volume is ${\sim}$19M\,m$^{3}$. The arm profile (dashed line) shows the elevation the arms of CE would need to be at to get Euclidean flat arms; the difference between the elevation (solid line) and the arm profile relate to the amount of earthworks needed at that location.}
    \label{fig:elevations}
\end{figure}

Laser beams travel in nearly straight lines, so Cosmic Explorer requires nearly straight beam tubes.
Most of the Earth's surface, however, does not follow this shape; ignoring terrain such as mountains and valleys, the Earth's surface is approximately spherical. 
For a spherical Earth model, the sagitta $s$ of the arc defined by a chord of length $L_{\rm arm}$, i.e., the amount of rise of the land above the center of a distance $L_{\rm arm}$ whose endpoints are on the surface, is defined by $s = R_{\Earth} - \sqrt{R_{\Earth}^2 - (L_{\rm arm} /2)^2}$, where $R_{\Earth}=6371$\,km is the mean radius of the Earth. While the sagitta is only 0.3\,m for the existing $L=4$\,km LIGO observatories, it rises to $8$\,m for $L_{\rm arm}=20$\,km and $31$\,m for $L_{\rm arm}=40$\,km. For CE40, the earthworks (cut, fill, bridges, and tunnels) required to create straight beam tubes at such a location would be enormous. 
Conversely, the cost of the earthworks will be least for land that is Euclidean flat (bowl-shaped in elevation), so that the land matches the path of the laser beam.
Figure~\ref{fig:elevations} illustrates this for a nearly constant-in-elevation site and a nearly Euclidean-flat site.

Building CE in locations that are nearly Euclidean flat will thus greatly reduce the amount of excavation and changes to the land required for its construction. 
For the locations with the most favorable elevation, constructing flat and level 40\,km-long arms will still require earthworks to level any deviations in the land from Euclidean flatness.

\begin{figure}
\centering
\includegraphics[width=0.75\textwidth]{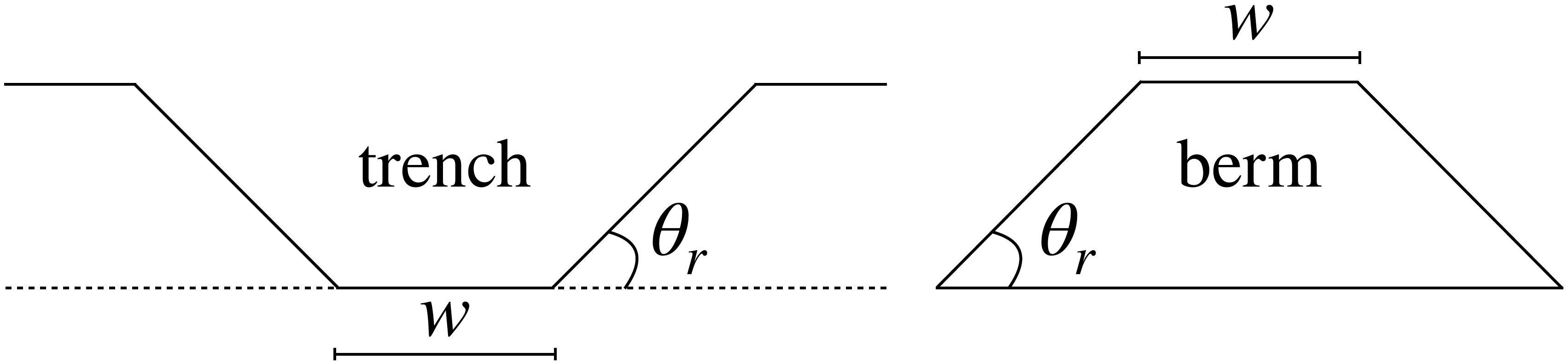}
    \caption{Illustration of trenches and berms needed to build straight arms for CE. The width $w$ is taken to be $w=$18\,m, and the angle of repose of both the trenches and berms is $\theta_{r}=45^{\degree}$.}
    \label{fig:trenchBerm} 
\end{figure}

We estimate the cost of earthmoving to bring the detector's footprint to Euclidean flat (``elevation cost'') from the volume of material that would need to be excavated or built up (cut and fill), with a cap on this cost where digging a tunnel would be less expensive than digging a trench. Specifically, we estimate the cost in terms of
\begin{equation}\label{eq:cutfill}
\Delta E = \left(V_{\rm cut} + V_{\rm fill} + |V_{\rm cut} -V_{\rm fill}|\right).
\end{equation}
The $V_{\rm cut} + V_{\rm fill}$ term corresponds to the amount of earth moved on the site, while the absolute value term corresponds to the amount of excess earth that must be brought to or removed from the site. 
For each trial site, the arm tilt chosen is the one that minimizes the excavation volume.
The tilt assigned to a given location is thus an outcome of the fit to minimize elevation cost.
Tilt is defined along the two arms, the x-arm and the y-arm, and further described in section~\ref{sec:tiltcost}. 
For pixels along the arm, we estimate cut and fill in terms of earthmoving required to build a trench or berm (figure~\ref{fig:trenchBerm}) with width $w$ and angle of repose $\theta_r$. (The angle of repose is the steepest angle with respect to a flat base at which a material can be piled along a slope without sliding down.) 
We note that this is only one model for potential earthworks; models including other factors, such as retaining walls, would be explored in a more detailed civil engineering costing. 

In CELS, we currently set the berm/trench width to $w=18.0$\,m, driven by the requirement of two 12-foot travel lanes and shoulders and one 14 ft (4m) beam tube enclosure, and we estimate the angle of repose at $\theta_r=45^{\circ}$, which is the value for gravel and several other typical ground materials. 
We note that trenches and berms are simplified representations of earthworks that could be used to provide a level CE beamline; in practice, civil engineering solutions such as tunnels, viaducts, or reinforced walls could be used. 

\begin{figure}
\centering
\includegraphics[width=.9\textwidth]{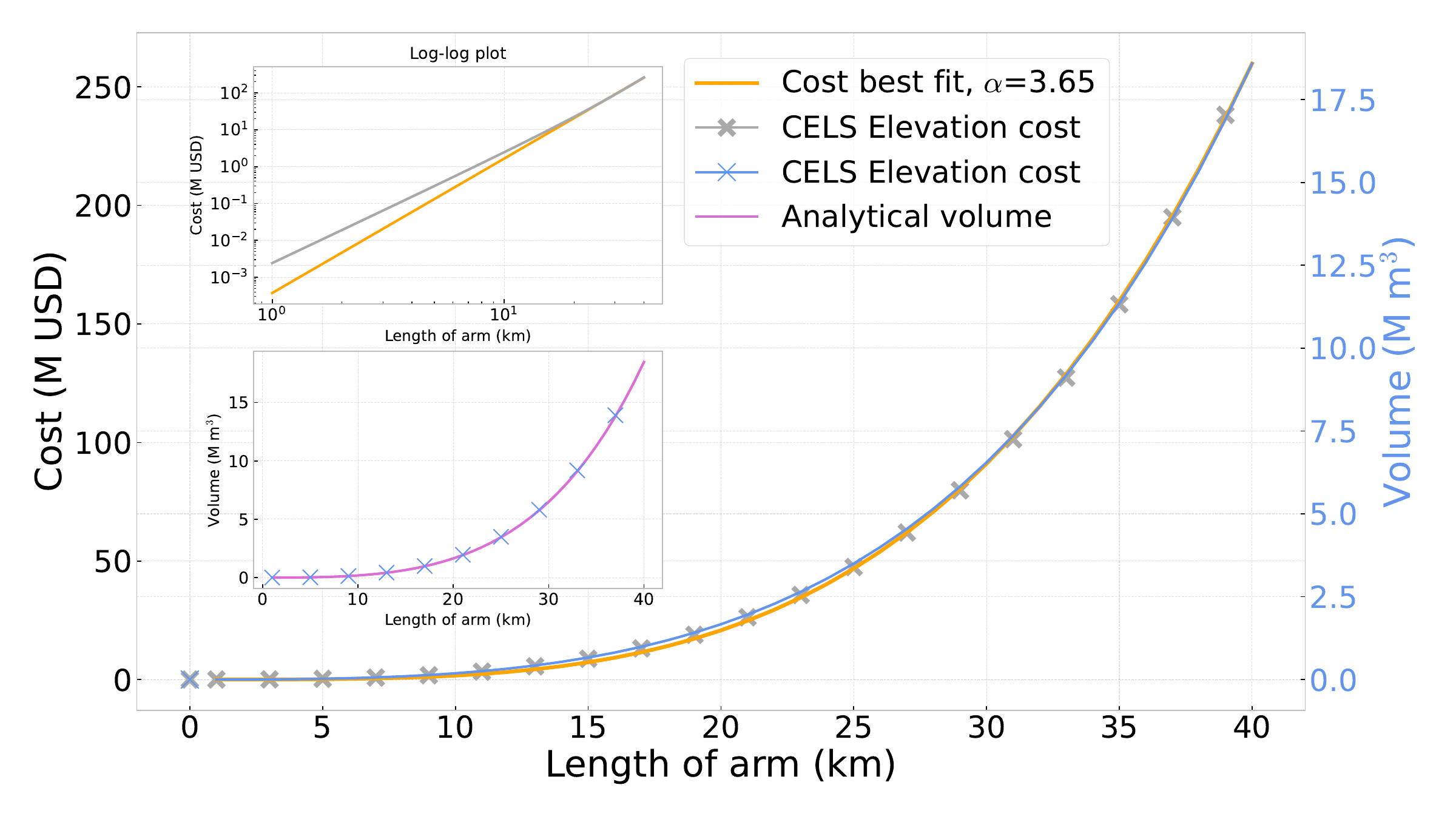}
    \caption{Plot showing how the cost and volume of earth moved scale versus arm length for a zero-elevation, spherical Earth model. This calculation uses CELS to score detectors of different lengths for terrain that everywhere has constant elevation. The calculation used an 18\,m berm and trench width. With a power law best fit and within the limits where $4\rm{\,km}\leq L_{\rm arm}\leq 40\rm{\,km}$, $\text{Cost} \propto L_{\rm arm}^{3.65}$. We note that CELS costs are not purely geometrical, because digging costs are capped to a fixed tunneling cost past a certain digging volume. However, in a zero-elevation case with 40\,km arms and a trench width of 18\,m, the digging depth never reaches the minimum tunnel depth.}
    \label{fig:LvsCost} 
\end{figure}

We represent the detector as a set of sample points at evenly spaced locations along the arms; the resolution $\Delta l$ of these points is a user-defined argument (taken to be $\Delta l = 100~{\rm m}$ in this paper unless stated otherwise). The earthworks volume, $V_k$, for each step $k$ along the arm is then
\begin{equation}
V_k  = h_k \Delta l (w + h_k \cot{\theta_r}),
\end{equation}
where $h_k$ is the depth at the $k$-th $\Delta l$-separated position along the arm, with negative values for cut (digging a trench) and positive values for fill (building a berm). 
The sum and difference in Eq.~\ref{eq:cutfill}, $V_{\rm cut}$ and $V_{\rm fill}$, are then given by

\begin{equation}\label{eq:numerator}
V_{\rm cut} + V_{\rm fill} = \sum_{k}|V_{k}|
\quad\text{  and  }\quad
V_{\rm cut} - V_{\rm fill} = \left|\sum_{k}V_{k}\right|.
\end{equation}

For a given $\Delta l$ position, we set a cap in cost for excavation depth only (i.e., not for building up berms), such that if the excavation volume for that trench is more than 500.0 m$^3$/m,
it remains fixed at that value. We set this cap under the assumption that for any deeper depth, building a tunnel would cost less than digging a trench of that depth.
The area of a trench with $\theta_{r}=45^{\degree}$ is
$h_{k} \times(|h_{k}| + w)$, so for $w=18.0$\,m, a cap of 500.0\,m$^3$/m corresponds to a height of  $h_{k} = -15.1$\,m. 

We then estimate the elevation cost $C_{\rm ele}$, which is the cost to level an L-shaped portion of earth to Euclidean flatness, in millions of USD as
\begin{equation}\label{eqn:Cele} 
    C_{\mathrm{ele}} = \frac{V_{\mathrm{cut}} + V_{\mathrm{fill}} + \left|V_{\mathrm{cut}}-V_{\mathrm{fill}}\right|}{10^6 \mathrm{m}^3}\times\$_{\mathrm{ave}},
\end{equation}
where volumes are given in units of ${\rm m}^3$, the terms in the numerator are given by Eq.~\ref{eq:numerator}, and where we estimate the cost of moving one cubic meter of earth as $\$_{\mathrm{ave}}=\$13$. This represents a rough average of estimates ranging from inexpensive excavation of topsoil (\$8/$\mathrm{m}^3$), sand and clay (\$13/$\mathrm{m}^3$) and gravel (\$22/$\mathrm{m}^3$), to expensive removal of wet soils and hard rock (\$100/$\mathrm{m}^3$). This estimate could be refined in future work by taking into account the geology of a given site. The factor of $10^6 $\,m$^3$ scales the overall costs to units of 1M USD. 
This is a coarse estimate and does not include the distance any material would need to be hauled. We also note that excavated material is greater in volume after removal; for this analysis, we ignore this effect on the resulting volume of moved material. Figure~\ref{fig:LvsCost} shows how CELS computes $C_{\rm ele}$ and the best fit to these scores for $L_{\rm arm}$ (m) vs cost (in millions of USD).

It is also possible to derive an analytical expression to get the volume $V$. As shown in Ref.~\cite{Schiettekatte_2026}, assuming a spherical Earth of radius $R_{\Earth}$ and $L_\mathrm{arm}\ll R_{\Earth}$, the volume of earth to displace in order to form berms or trenches for a single arm, as illustrated in figure~\ref{fig:trenchBerm}, with $\theta_{r} = 45{^\circ}$ is given by

\begin{equation}\label{eq:exactVol}
V = \frac{8}{3}w z_0^{3/2}\sqrt{2R_{\Earth}} + 2z_0(z_0 - w)\frac{L_\mathrm{arm}}{2}  - \frac{2z_0 - w}{3R_{\Earth}}\left( \frac{L_\mathrm{arm}}{2} \right)^{3} + \frac{1}{10 R_{\Earth}^{2}}\left( \frac{L_\mathrm{arm}}{2} \right)^{5},
\end{equation}
\noindent where $z_0 > 0$ is the depth of the trench at the middle of an arm, i.e., $-h_k$ at $k = L_\mathrm{arm}/2$. It follows a polynomial with odd-power terms up to $L_\mathrm{arm}^{5}$, highlighting the strong dependence of the volume on $L_{\mathrm{arm}}$, hence costs. In the special case where the ends of the arm are at-grade, $z_0 = L_\mathrm{arm}^{2}/8R_{\Earth}$ (which is the sagitta) and Eq.~\ref{eq:exactVol} reduces to

\begin{equation}\label{eq:VolAtGrade}
V = \left( \frac{1}{15}\left( \frac{L_\mathrm{arm}}{R_{\Earth}} \right)^{2} + \frac{2w}{3R_{\Earth}} \right)\left( \frac{L_\mathrm{arm}}{2} \right)^{3},
\end{equation}

\noindent as indicated in Ref.~\cite{10.1063/5.0242016}. One can also find the optimal value of $z_0$ to minimize Eq.~\ref{eq:exactVol}. The resulting expression, detailed in Ref.~\cite{Schiettekatte_2026}, depends on $L_\mathrm{arm}$, but varies smoothly between $L_\mathrm{arm}^{2}/32 R_{\Earth}$ for short arms and $L_\mathrm{arm}^{2}/24 R_{\Earth}$ for long arms. Eq.~\ref{eq:exactVol} using this optimal $z_0(L_\mathrm{arm})$ (or actually $2V$ for two arms) matches CELS calculations, as seen in figure~\ref{fig:LvsCost}.

While this confirms that CELS gives correct results in this case, it is difficult to see from Eq.~\ref{eq:exactVol}, also considering the $z_0(L_\mathrm{arm})$ dependency, which terms dominate costs in different situations. For this reason, we fit a power law to the elevation costs, which helps capture the trend. For 40\,km arms, the best fit goes as $L_\mathrm{arm}^{3.65}$, which is also represented in figure~\ref{fig:LvsCost}. It fits accurately the costs (and, by extension, volume) above 30\,km, while it departs significantly below 10\,km. Yet it shows that the actual trend sits somewhere between the two highest powers of the polynomial of Eq.~\ref{eq:exactVol} near $L_\mathrm{arm} = 40$\,km. 

This strong dependence of volume of earth moved on arm length makes elevation a primary cost driver for CE. For example, in the spherical Earth approximation, the total volume of excavated material for CE40 is equal to ${\sim}19$M\,m$^3$, or the equivalent of ${\sim}2$ million excavation vehicle loads of moved material. This is prohibitive from a cost perspective unless (which seems unlikely) extreme economies of scale could greatly reduce the cost, leading to, for example, significantly lowering smaller tunnel or trestle bridge costs.

This underscores the benefits (lower cost, less change to the existing land) of identifying sites that are nearly Euclidean flat, rather than nearly constant in elevation. There are in fact locations in the conterminous U.S. that are nearly Euclidean flat, which corresponds to a gentle valley in elevation and often occur in soft sedimentary basins. Figure~\ref{fig:elevations} shows a comparison between two fiducial locations: a nearly zero elevation location in Georgia and a nearly Euclidean flat location near Monahans, TX. The excavation volumes of these two sites are approximately 19M\,m$^3$ and 4M\,m$^3$, respectively.

CELS currently minimizes earthworks costs by adjusting the elevation and tilt of the detector arms. Sometimes, this leads to situations where the corner and end stations would not be near grade. 
It is preferable for CE's science outcomes to not have end stations built on a viaduct ($> 20$\,ft above grade) or tall structure that could have increased motion. Buried stations could have reduced vibration from seismic and wind. From a workforce and accessibility standpoint, it is preferable to have the stations at-grade. 
These considerations are in tension with the fact that at-grade stations will tend to increase earthworks costs---especially the likelihood of building costly tunnels. Future implementations of CELS will search for the best places across the conterminous United States to build a CE with at-grade or buried end stations.

The theoretical elevation cost described in this section is likely not better than ``order of magnitude" accuracy. To improve the accuracy and calibration of excavation cost calculations with CELS, we expect to contract with a civil engineering firm to reassess this estimate. Further work is underway to modify the costs based on different soil and rock types. 

\subsection{Impact on Science}\label{sec:science}

Flexibility in the layout and placement of a site, e.g., allowing the length and opening angle of the arms to vary, allows for a tradeoff between sensitivity, cost, and siting flexibility. 
Some factors related to the overall site layout---including tilt, deviations from a 90$^{\degree}$ arm opening angle, and shorter arm lengths---will diminish the scientific returns for a given detector configuration with respect to an untilted, 90$^{\degree}$ opening angle, and $40$\,km (or $20$\,km) length. 
Since the excavation costs associated with preparing a detector footprint  scale with a high power of $L_{\rm arm}$ for locations that are not nearly Euclidean flat, detectors with shorter $L_{\rm arm}$ will always cost less, making cost alone a poor ranking statistic. Alternatively, a ``science per dollar" metric takes into account the effects both on scientific outcomes and on costs for a given detector layout. 
We did not use a science per dollar metric for the national searches presented here, but we do anticipate such a metric becoming increasingly useful as CE moves forward toward more focused, local analyses of potential sites. In section~\ref{subsubsec:science}, we describe such a science per dollar metric. Then, in section~\ref{sec:tiltcost}, we describe how CELS already considers one factor impacting science: tilt.

\subsubsection{Arm length and opening angle}\label{subsubsec:science}
A two-armed laser interferometer gravitational-wave observatory, such as CE, has an amplitude sensitivity to a given astrophysical source that scales with the quantity $L_{\rm arm}\sin{\theta_{\rm oa}}$, where $L_{\rm arm}$ is the length of the arm and $\theta_{\rm oa}$ is the opening angle between the arms.
Following the approach outlined in Ref.~\cite{Read_2023}, the effect of this scaling can be modeled as a power law, $(L_{\rm{new}}\sin{\theta_{\rm{oa,new}}})^{\gamma}$, characterizing the change in an observational metric from adding a detector with length $L_{\rm arm} = L_{\rm{new}}$ and opening angle $\theta_{\rm oa} =\theta_{\rm{oa,new}}$ to an existing gravitational-wave detector network. 
Specifically, in Ref.~\cite{Read_2023}, the following metrics are considered: i) signal-to-noise ratio (SNR) of a fixed source, ii) the number of sources with a given SNR, and iii) the distances (and corresponding time-sensitive volumes) to which binary black holes and binary neutron stars can be observed. For a gravitational-wave detector of approximately 10 times the sensitivity of current detectors, added to a network of comparable strength, these metrics scale with powers between 1.6 and 2.2, motivating the choice of a fiducial $\gamma=2$. A combination of cosmological factors and the relative impact of the existing network reduce this power-law scaling compared to the $\gamma=3$ scaling expected for a single observatory where the observable volume of space-time scales as the cube of the amplitude sensitivity.

The CE site search is currently considering arm opening angles of 65\textendash115$^{\degree}$ and arm length reductions of 4\,km (for CE40) or 2\,km (for CE20). These ranges correspond to a maximum ${\sim}10$\% strain amplitude reduction with regards to the ideal detector configuration. 
Following the discussion above, the scientific signal benefit of a detector with an opening angle $\theta_{\rm oa}$ and an arm length $L_{\rm arm}$, normalized to 1 with respect to one with $\theta_{\rm oa}=90^{\degree}$ and 40\,km, respectively, can then be estimated as
$\mathcal{S} = \left(\left(L_{\rm arm}\sin{\theta_{\rm oa}}\right)/40\,\rm{km}\right) ^{2}$.
This quantity can then be divided by the project costs (e.g., as estimated by CELS) for the same location, arm length $L_{\rm arm}$, and opening angle $\theta_{\rm oa}$, to yield an estimate of science per dollar.

\subsubsection{Tilt}
\label{sec:tiltcost}

The elevation cost considerations discussed in subsection~\ref{ss:elevation} favor Euclidean-flat (bowl-shaped in elevation) topographies, because they reduce earthmoving costs. To identify the optimal detector configuration that minimizes earthmoving costs, CELS permits the end and corner stations to have different elevations, i.e., for the detector arms to be tilted with respect to the local gravitational field. For instance, CELS would estimate an elevation cost of zero for a hypothetical topography that is perfectly Euclidean flat yet on a 45-degree slope.

\begin{figure}[h]
\centering
\includegraphics[page=6,width=0.49\textwidth]{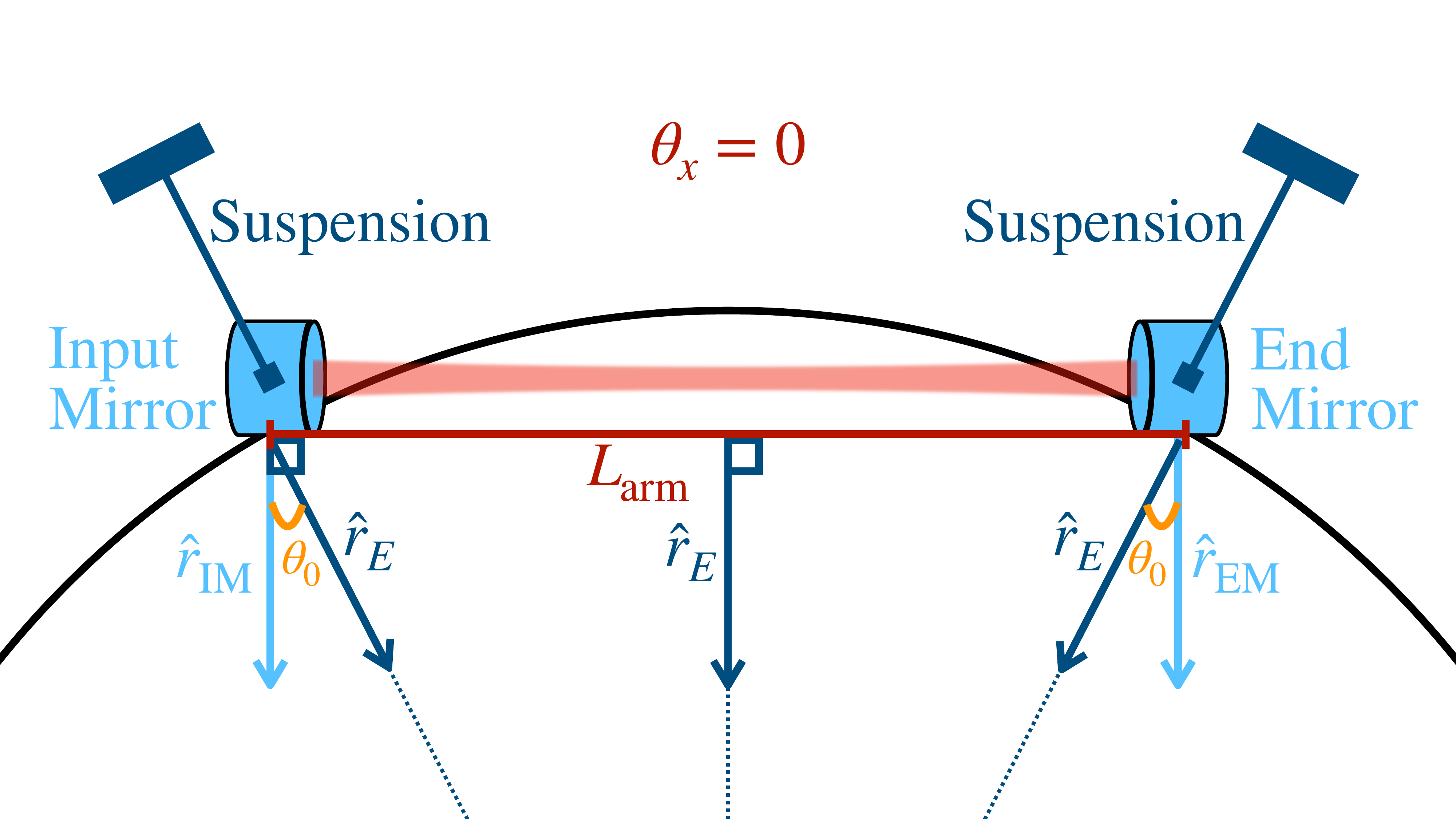}
\includegraphics[page=7,width=0.49\textwidth]{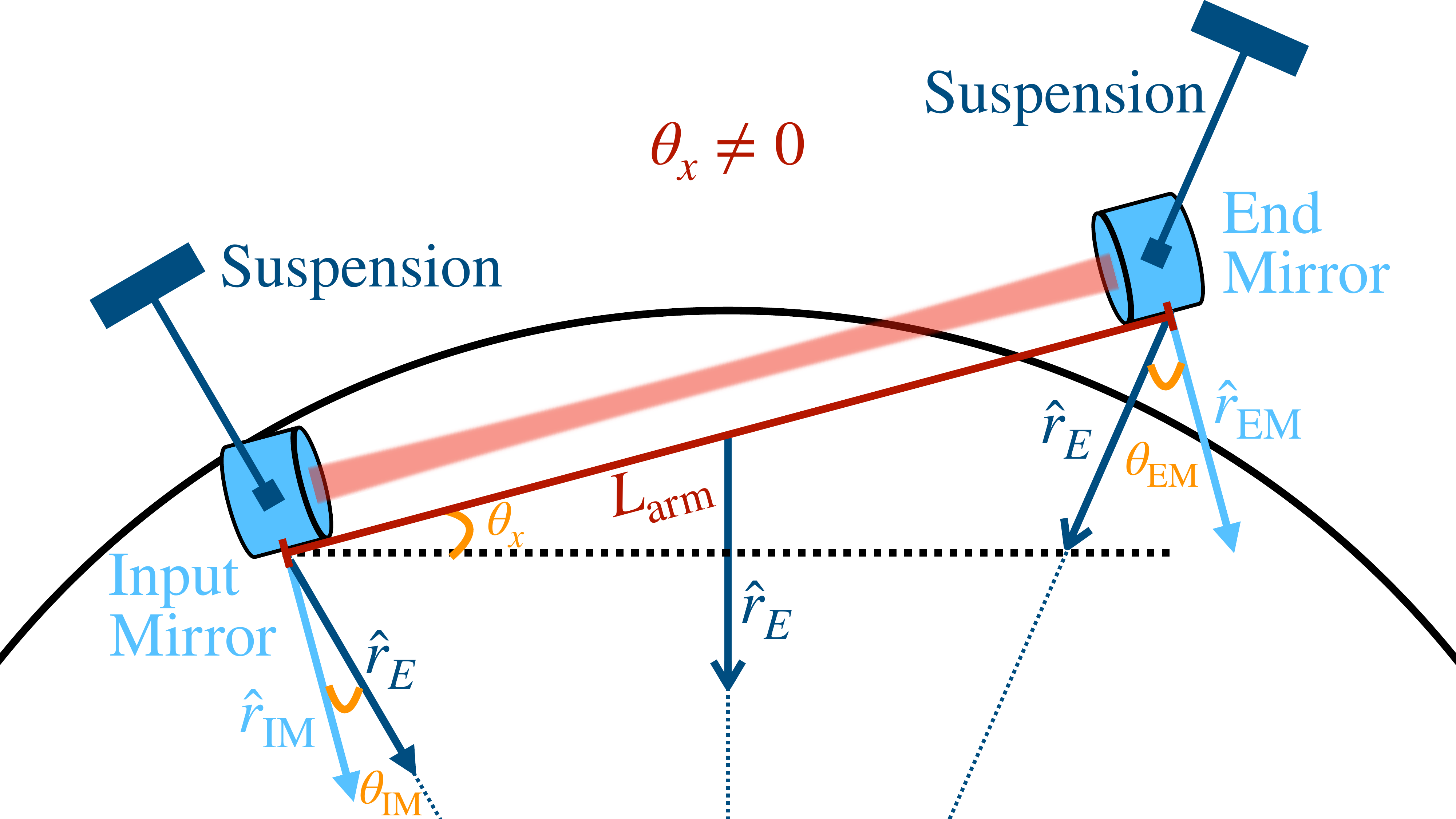}
    \caption{Illustrations of an x-arm of CE, showing the mirror suspensions for the input mirror at the corner station, and the end mirror at the x-end station. The mirrors are separated by length $L_{\rm arm}$ and they form an optical cavity that reflects the laser beam back and forth. The black curve indicates a spherical Earth, which corresponds to zero elevation everywhere. The length of CE's arm is greatly exaggerated here, and in reality is about 1/1000th the Earth's circumference. 
    \textit{Left:} An x-arm of CE with no tilt ($\theta_x=0$). Since the mirror suspensions follow the local gravitational field, each mirror must be pitched inward by $\theta_{0}$ (in reality, only 3.1\,mrad) to reflect the laser toward the other mirror.  
    \emph{Right:} Here the x-arm has a tilt, $\theta_x\neq0$. The input mirror and end mirrors now have pitch angles with respect to the local gravity direction of $\theta_{\rm IM}=\theta_0-\theta_x$ and $\theta_{\rm IM}=\theta_0+\theta_x$, respectively. A tilted arm will always couple more vertical vibrations into CE's gravitational-wave readout.}
    \label{fig:tilt} 
\end{figure}

Such a tilted site would not be ideal for realizing CE's science goals. This is because CE, like LIGO, will have mirrors that are suspended at the bottom of multiple-tiered pendulums that provide isolation from seismic noise and other noise sources~\cite{Aston_2012}.  Optics suspended from a pendulum align with the local direction of effective gravity at their location, hanging perpendicular to Earth's surface (assuming a spherical Earth). Since the laser beams in the detector arms travel in essentially straight lines, and since creating an optical cavity in the arms requires both mirror surfaces to be perpendicular to the laser beam, the mirrors must be statically pitched forward to compensate for the angle of the pendulum with regards to the laser. 
As illustrated in figure~\ref{fig:tilt}, this pitch angle for an arm with zero tilt (i.e., where the vertex and end are at the same elevation) is (adopting a spherical-Earth approximation) $\theta_{0} = L_{\text{arm}}/2R_{\Earth} \approx 3.1\text{\,mrad}$, where $R_{\Earth}$ is the average radius of the Earth and $L_{\text{arm}} = 40$\,km is the arm length)~\cite{Kuns_2020}. 
If an arm is instead tilted, say the x-arm at $\theta_x$, one of the mirrors will have a pitch angle with respect to the local gravity of $\theta_0-\theta_x$ while the other will have $\theta_0+\theta_x$.
Pitching the mirrors with respect to local gravity causes vertical motions of the pendulums to be coupled more strongly into horizontal motions of the mirror---the degree of freedom for gravitational-wave readout. 
This is important because, in gravitational-wave detectors like Cosmic Explorer, the vibration isolation in the vertical direction tends to be less effective than the horizontal isolation, and even a small vertical to horizontal coupling can add significant noise. 
For example, the power spectral density of intrinsic vibrations in a given suspension caused by thermal noise would be coupled by the square of the tilt angle:
$S^{\text{tot}}(\omega) = S^{\text{hor}} + \theta^{2}S^{\text{vert}}(\omega)$,
where $\theta$ is the pitch angle with respect to local gravity for the corresponding mirror~\cite{PhysRevD.108.123009}. Using this type of coupling as a model and considering the uncorrelated contributions of the four main mirrors in CE's two arms, we note that any nonzero tilt will add noise to CE's gravitational-wave measurements (increasing the amplitude spectral density quadratically for $\theta_x\lesssim\theta_0$ then linearly for $\theta_x>\theta_0$), which would degrade the science CE could achieve~\cite{celsTilt}.

\begin{figure}
\centering
\includegraphics[width=\textwidth]{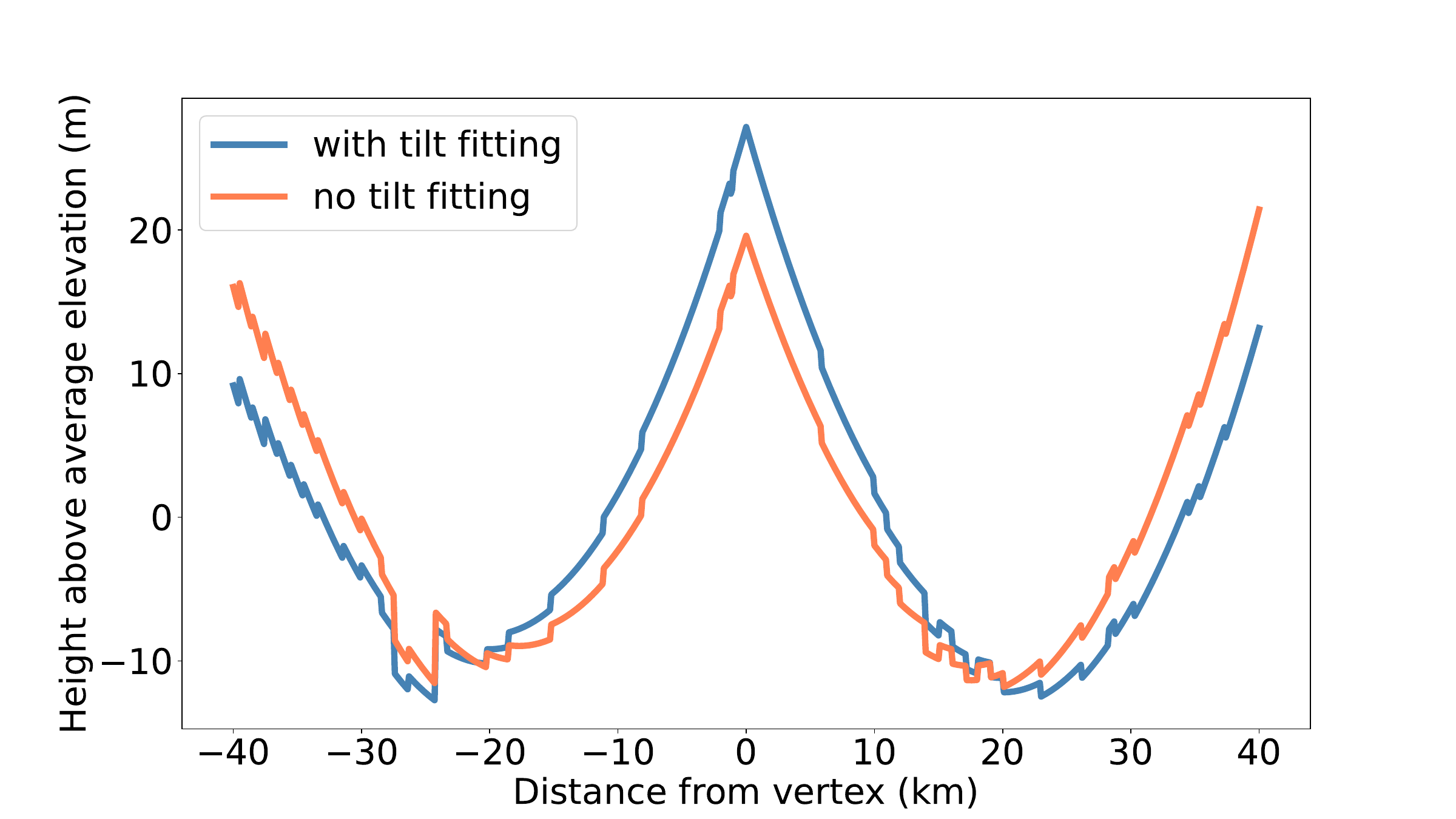}
    \caption{For a fiducial location, height of the earthworks needed to build Euclidean-flat beam tubes when arm tilt is fitted to minimize the total earthmoving volume (blue) and when arm tilt is kept to be 0 (orange). When the arm tilt is fitted to minimize the elevation costs, the earthmoving volume is ${\sim}18$ million cubic meters. When the arm tilt is not fitted, that volume is ${\sim}20$ million cubic meters. This represents a difference in CELS elevation cost of ${\sim}40$M USD.}
    \label{fig:tilt_fit} 
\end{figure}

We currently consider tilt in CELS as a ``cost". 
Arm tilt can be expressed as $\theta_x$ and $\theta_y$, in radians (m/m) for the x- and y-arms of the detector, respectively.
We note that while $\theta_x$ and $\theta_y$ could, in principle, be zeroed at any location, this zeroing would typically come with an increase in earthworks cost. Because earthworks are a driving cost for CE, CELS optimizes the tilt to minimize earthworks cost $C_{\rm{ele}}$ associated with the topography of the site. However, to prevent CELS from selecting configurations that are close to Euclidean flat but with a steep tilt, CELS also computes a ``cost" $C_{\rm tilt}$ associated with the tilt of the detector arms in CELS. Specifically,
\begin{equation}
C_{\rm tilt} = 10 \rm{M\,\,USD} \left[\left(\frac{\theta_{x}}{\theta_{0}}\right)^{2} + \left(\frac{\theta_{y}}{\theta_{0}}\right)^{2}\right],
\end{equation}
with $\theta_{x}$ and $\theta_{y}$ the tilts of the x- and y-arms, respectively. The factor of 10 M USD represents a rough attempt to quantify the added ``cost" associated with the noise and control complexity added by the tilt.
We note that this ``cost" scales quadratically with tilt, penalizing highly tilted arms and in rough agreement with the quadratic scaling with amplitude sensitivity rule of thumb for science metrics discussed above in section~\ref{subsubsec:science}.
Future work will seek to incorporate tilt considerations in CELS's science per dollar metrics, rather than treating it as a monetary cost. Currently, there is no maximum allowable tilt, but future work to better characterize its impact on CE's science outputs will allow us to put an upper allowable cap on tilt.

\section{Results}\label{sec:results}

We have used CELS to search the conterminous United States for cost-favorable locations for CE. Previous work~\cite{Datrier:2025wjs} presented results from a national search using the following three inputs to the estimated cost: land cover, elevation, and tilt. Here we present results from two new national searches, in which we i) incorporated additional inputs (land acquisition costs and major road crossings) to the estimated cost and ii) improved the estimates for land cover cost. For a discussion of each input used for the results presented here, see section~\ref{ss:datasources}. The two national searches (one for CE40 and one for CE20) used a 1\,km grid over the conterminous United States. The search estimated the cost to construct a detector with a vertex at each grid point with 36 different rotations of the arms about the vertex, and a 90$^\degree$ opening angle between the detector arms. Because each cost estimate in this search corresponds to a 90$^\degree$ opening angle and
either a 40\,km or 20\,km arm length, the scientific signal benefit as described in section~\ref{subsubsec:science} is always 1. Future, more localized searches, will consider other arm lengths and opening angles and thus other scientific signal benefit factors.

\subsection{Results for a 40\,km Cosmic Explorer}

\begin{figure}
\centering
\includegraphics[width=.9\textwidth]{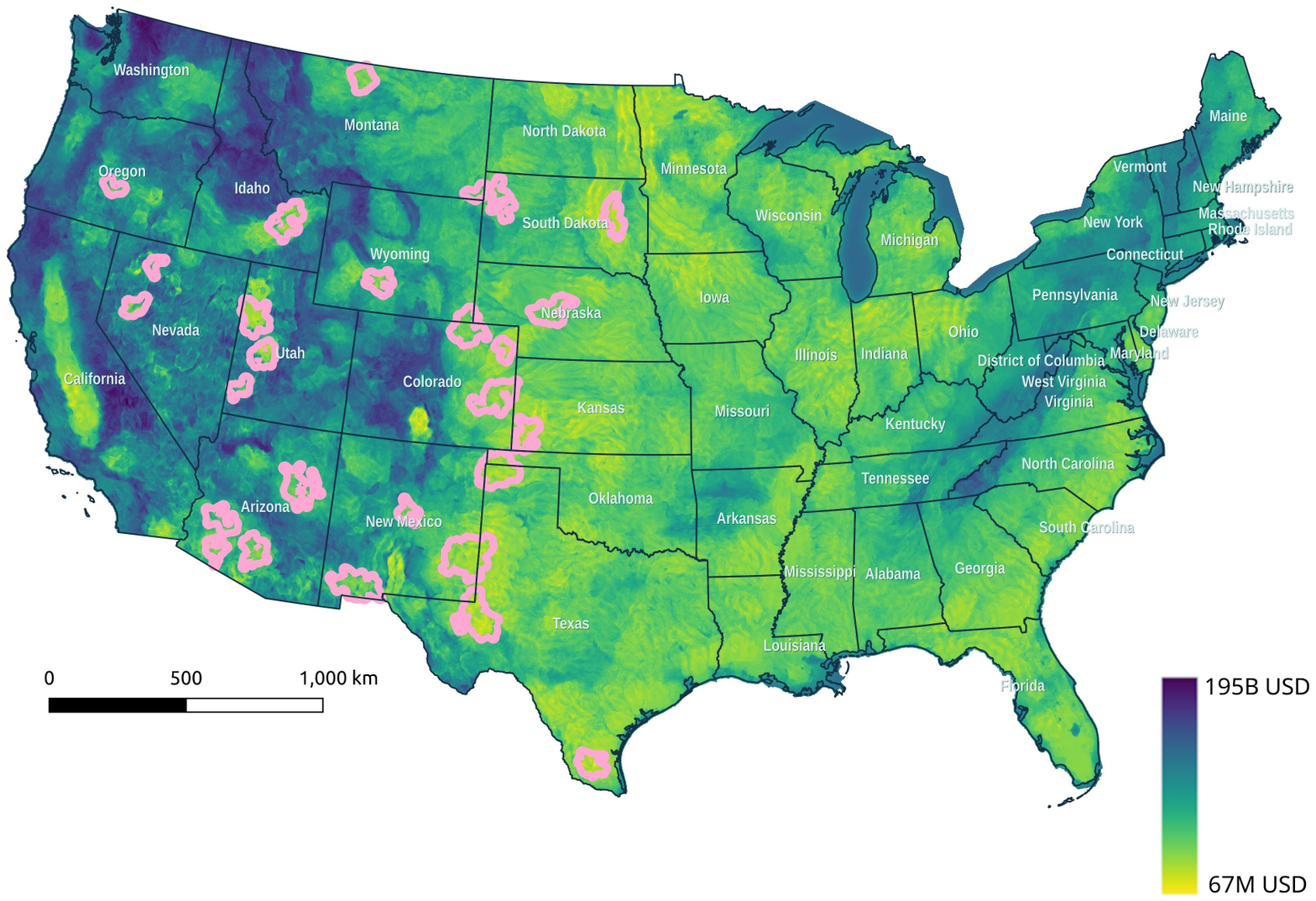}
\hspace{0.1in}
\caption{Results of the national search for CE40 locations. Each pixel represents the estimated minimum cost over 36 evenly spaced rotations for a given location of the detector vertex (corner station). Yellow represents areas with the lowest associated cost estimates, while purple represents areas with the highest cost estimates. The color scale is logarithmic. The pink outlines show regions containing cost-favorable potential locations for CE40, corresponding to 27 of the 29 long-listed sites currently under consideration. The two locations that are not shown on the map were identified as areas of interest by the National Suitability Analysis only~\cite{Bristol_2026}. 
Note that the pink outlines do not necessarily reflect the regions with the absolute lowest cost estimates identified in this national search, but instead show the lowest-cost options in and near these long-listed areas. The CE site evaluation team developed the long list with input from our previous CELS national search results, the National Suitability Analysis, and other efforts to identify favorable potential sites (see section~\ref{sec:intro}). \label{fig:40km}}
\end{figure}

\begin{figure}
\centering
\includegraphics[width=.9\textwidth]{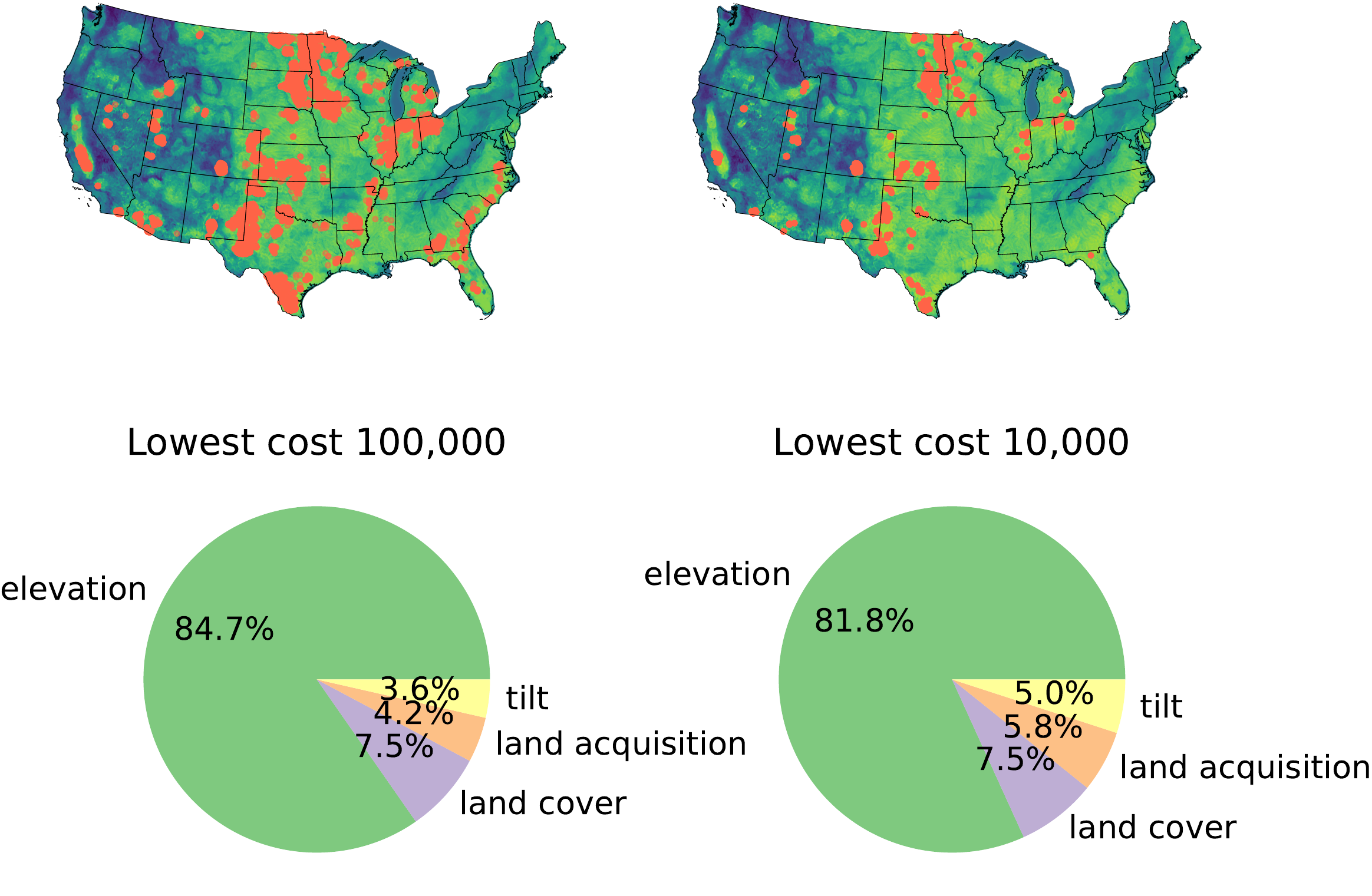}
\hspace{0.1in}
\caption{Contributions to cost estimates of the 100,000 and 10,000 CE40 configurations with the lowest cost returned by the CELS national search shown in figure~\ref{fig:40km}. The pie charts omit road crossing costs, because they are negligible
  ($<1\%$ of the total cost estimate) in this subset of the data. (Road crossings tend to be cost prohibitive, and thus tend not to appear in sites with the most favorable cost estimates.) For this subset, elevation (i.e., earthworks to make a site Euclidean-flat) represents more than 80\% of the cost estimates on average. \label{fig:piechart_costs}}
\end{figure}

We have prioritized searches for CE40 observatories, in line with the recommendations for the global gravitational-wave detector network outlined in~\cite{KalogeraReport}, and recognizing that potential CE40 sites would also accommodate CE20 observatories.
Figure~\ref{fig:40km} presents the results of our new 40\,km national search. The figure shows a bitmap, where each pixel's color corresponds to the lowest cost estimate out of all detector rotations with a vertex at that pixel's location. The color scale is logarithmic, with yellow representing the lowest cost estimates and purple representing highest cost estimates. For CE40, the lowest cost estimate is ${\sim}55$M USD.
Elevation (i.e., earthworks to make site topography  Euclidean flat) drives most of this estimated cost; figure~\ref{fig:piechart_costs} shows the contributions of each factor contributing to the total cost estimate for the 100,000 and 10,000 lowest cost sites. The figure shows that earthworks costs related to terrain make up more than 80\% of the total site preparation cost estimate.

\subsection{Results for a 20\,km Cosmic Explorer}
\begin{figure}
\centering
\includegraphics[width=.9\textwidth]{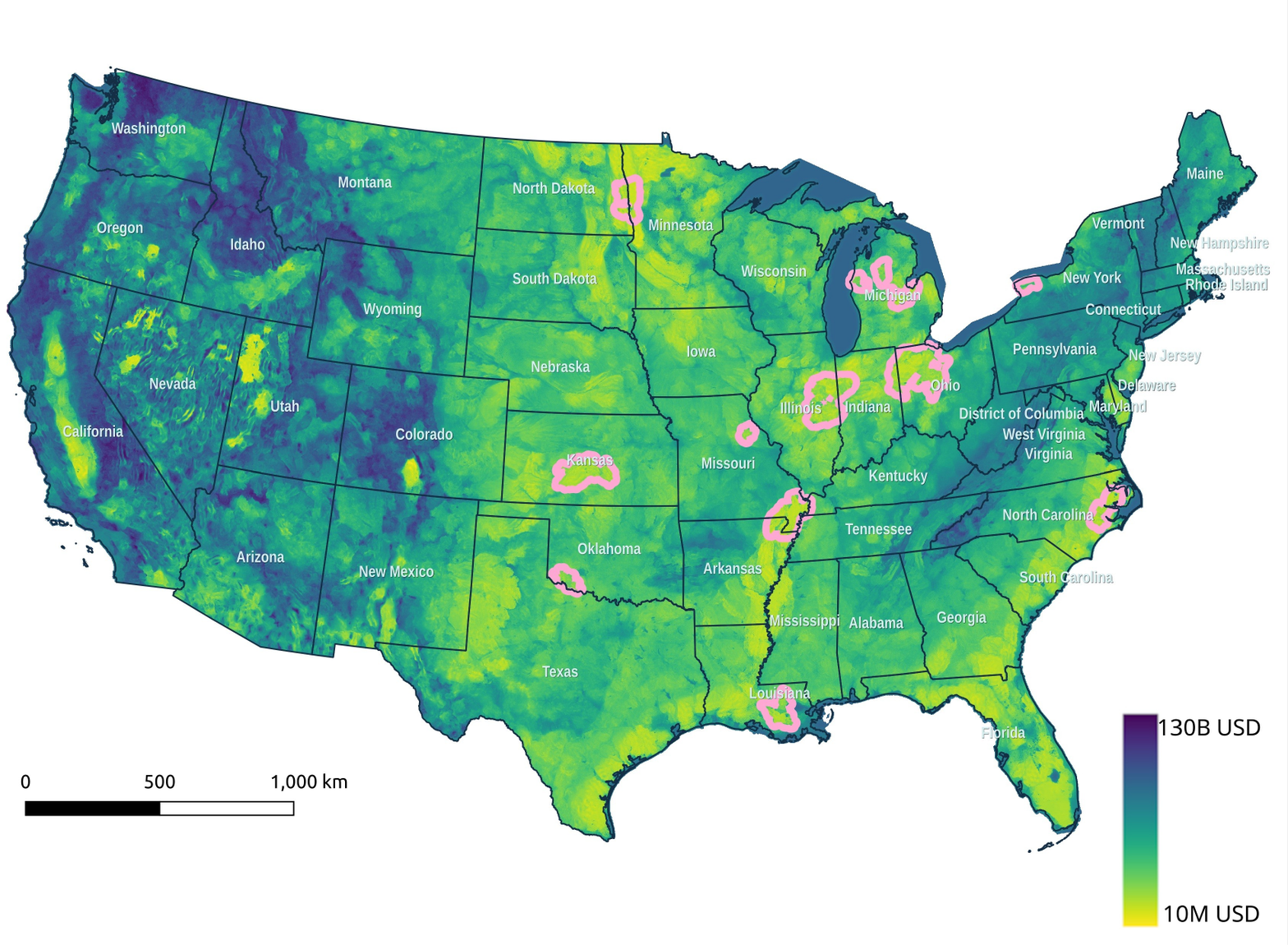}
\hspace{0.1in}
\caption{Results of the national search for CE20 locations. Each pixel represents the estimated minimum cost over 36 evenly spaced rotations for a given location of the detector vertex (corner station). Yellow represents areas with the lowest associated cost estimates, while purple represents areas with the highest cost estimates. The color scale is logarithmic. The pink outlines show regions containing cost-favorable potential locations for CE20 in and near areas of interest identified by the National Suitability Analysis and CELS. As discussed in the text, we favor locations in the eastern United States for CE20.
  Note that the pink outlines do not necessarily reflect the regions with the absolute lowest cost estimates identified in this national search.\label{fig:20km}}
\end{figure}

Figure~\ref{fig:20km} shows the new national search results for CE20; the lowest cost estimate is ${\sim}8$M USD.
We favor locations in the eastern United States for CE20
for the following reasons.
First, the CE20 search identifies options for cost-favorable potential sites in the eastern United States, while the CE40 search did not. Pairing a CE40 observatory in the western United States with a CE20 in the eastern United States leads to a longer baseline between the two detectors, which is important for localizing the sky locations of gravitational-wave sources. For binary black hole (BBH) localization, given a relative orientation (difference in the detectors' rotation angles) of $45^{\degree}$, a baseline of 2300--3300\,km is best for eliminating multi-modalities in the skymap~\cite{iacovelli2026closeevaluatingimpactbaseline}. The minimum baseline currently under consideration for a CE40--CE20 pair is 1000\,km~\cite{10.1063/5.0242016}. Because CE20 would not be built without CE40, and because most cost-favorable locations for CE40 are in the midwestern to western United States, a CE20 observatory in the eastern United States would best achieve a long baseline between CE20 and CE40.

\section{Conclusion and future work}\label{sec:conclusion}

The CELS code has enabled us to identify cost-favorable areas for potential CE sites through site clearing cost estimates.

The broader CE site evaluation study has identified an initial list of 29 suitable locations for a CE40 observatory, and a further 13 locations in the eastern U.S. that could accommodate a CE20 observatory. We emphasize that these are draft candidate locations; the list of potential locations might evolve as we continue analyzing and visiting locations of interest. The CE site evaluation team expects to complete an initial report to the NSF on these locations by fall 2026. The NSF will determine the final site selection process.

In the future, we will use CELS work to carry out localized remote evaluations that focus on locations of interest. Updates to CELS will incorporate more accurate cost estimates and science factors, following consultations with engineers and architects, and will include more realistic earthmoving costs, earthworks costs for different types of rock, soil, and depth to bedrock, and more realistic tunnel and bridge costs.
We will more deeply integrate future CELS analyses with results from the National Suitability Analysis pipeline.

Future searches will also incorporate the scientific impact of the baseline and relative orientation between a CE40 and a CE20, in order to rank the best pairs of detectors based on several science metrics. For example, a 45$^{\degree}$ relative angle between CE20 and CE40 would result in high performance for localizing gravitational-wave observations but would lead to a reduced sensitivity to the stochastic gravitational-wave background~\cite{iacovelli2026closeevaluatingimpactbaseline, PhysRevD.110.122006, Romano2017}. Sky localization must also consider the locations of other observatories in the global next-generation gravitational-wave detector network. Besides baseline, orientation, detector arm length and opening angle, the corner and end station heights above or below grade can also affect CE's science goals (see discussion in section~\ref{ss:elevation}). Future searches will seek cost-favorable sites where CE's stations would be at, or slightly below grade.

The current national CELS results provide cost and positioning insight, helping to identify and evaluate cost-favorable sites for building Cosmic Explorer. 
This coarse, national search for locations for CE20 and CE40, together with the National Suitability Analysis results, will form the basis for future local and site-level analysis. These, along with site visits, relationship building, and instrumented assessments, will inform a long list and then a short list of potential locations for Cosmic Explorer.

\section*{Data Availability}

The results presented in this paper were obtained using CELS v.~2026.03.12. CELS is available at \url{https://gitlab.com/cosmic-explorer/cels/}. The data layers are publicly available data described in section~\ref{ss:datasources}. To obtain the results presented in this paper, the outputs from CELS v.~2026.03.12 were post-processed to account for inflation.
Final outputs in the form of GeoTIFFs and delimited text files are available on request.

\ack

We are grateful to colleagues in the Cosmic Explorer Project for discussions regarding observatory siting and costing and to Michael L. Dennis for his advice on geodesy.
We benefited from discussions with the integrated design firm SmithGroup, who conducted the 2026 Cosmic Explorer architectural and engineering feasibility study. This work would not have been possible without the initial work of Matt Evans, Kevin Kuns, and Evan Hall on the predecessor code to CELS, \verb|ifo-search|.
The Cosmic Explorer site evaluation effort is supported by NSF grants 2308985, 2308986, 2308987, 2308988, 2308989, and 2308990. The Cal State Fullerton authors were additionally supported by the U.S. National Science Foundation grant 2219109 and by Dan Black and Family and Nicholas and Lee Begovich. We used ChatGPT Deep Research for gathering references about and checking and guiding the rough cost estimates for bridges, tunnels, and excavation, and we used Cursor and Codex for some CELS code development.

\bibliography{bibli}

@misc{Read_2023,
    title={{Component Interferometer Length Scaling for Network Science Metrics}},
  author = {Read, J.},
  collaboration = {Cosmic Explorer},
  year = {2023},
  url = {https://dcc.cosmicexplorer.org/T2300001/public}
}

@misc{Schiettekatte_2026,
title = {{Analytical expressions for optimizing the digging/filling volume for installing long straight arms}},
author = {Schiettekatte, F.},
collaboration = {{Cosmic Explorer}},
url = {https://dcc.cosmicexplorer.org/CE-T2600009/public}

}

@article{Aston_2012,
doi = {10.1088/0264-9381/29/23/235004},
url = {https://doi.org/10.1088/0264-9381/29/23/235004},
year = {2012},
month = {oct},
publisher = {IOP Publishing},
volume = {29},
number = {23},
pages = {235004},
author = {Aston, S M and Barton, M A and Bell, A S and Beveridge, N and Bland, B and Brummitt, A J and Cagnoli, G and Cantley, C A and Carbone, L and Cumming, A V and Cunningham, L and Cutler, R M and Greenhalgh, R J S and Hammond, G D and Haughian, K and Hayler, T M and Heptonstall, A and Heefner, J and Hoyland, D and Hough, J and Jones, R and Kissel, J S and Kumar, R and Lockerbie, N A and Lodhia, D and Martin, I W and Murray, P G and O’Dell, J and Plissi, M V and Reid, S and Romie, J and Robertson, N A and Rowan, S and Shapiro, B and Speake, C C and Strain, K A and Tokmakov, K V and Torrie, C and van Veggel, A A and Vecchio, A and Wilmut, I},
title = {Update on quadruple suspension design for Advanced LIGO},
journal = {Classical and Quantum Gravity}
}

@misc{theligoscientificcollaboration2026gwtc50introductionversion50,
      title={GWTC-5.0: An Introduction to Version 5.0 of the Gravitational-Wave Transient Catalog}, 
      author={{The LIGO Scientific Collaboration} and {the Virgo Collaboration} and {the KAGRA Collaboration} and others },
      year={2026},
      eprint={2605.27223},
      archivePrefix={arXiv},
      primaryClass={gr-qc},
      url={https://arxiv.org/abs/2605.27223}, 
}

@Article{Romano2017,
author={Romano, Joseph D.
and Cornish, Neil. J.},
title={Detection methods for stochastic gravitational-wave backgrounds: a unified treatment},
journal={Living Reviews in Relativity},
year={2017},
month={Apr},
day={04},
volume={20},
number={1},
pages={2},
issn={1433-8351},
doi={10.1007/s41114-017-0004-1},
url={https://doi.org/10.1007/s41114-017-0004-1}
}

@article{PhysRevD.110.122006,
  title = {Next-generation global gravitational-wave detector network: Impact of detector orientation on compact binary coalescence and stochastic gravitational-wave background searches},
  author = {Ebersold, Michael and Regimbau, Tania and Christensen, Nelson},
  journal = {Phys. Rev. D},
  volume = {110},
  issue = {12},
  pages = {122006},
  numpages = {18},
  year = {2024},
  month = {Dec},
  publisher = {American Physical Society},
  doi = {10.1103/PhysRevD.110.122006},
  url = {https://link.aps.org/doi/10.1103/PhysRevD.110.122006}
}

@misc{ligoindiadcc,
title = "{LIGO-INDIA, Proposal of the Consortium for
INDian Initiative in Gravitational wave Observations (IndIGO)}",
author = {Bala Iyer and Tarun Souradeep and CS Unnikrishnan and Sanjeev Dhurandhar and Sendhil Raja and  
Anand Sengupta},
year = {2011},
url = {https://dcc.ligo.org/ligo-m1100296/public}
}

@misc{iacovelli2026closeevaluatingimpactbaseline,
      title={Not too close! Evaluating the impact of the baseline on the localization of binary black holes by next-generation gravitational-wave detectors}, 
      author={Francesco Iacovelli and Luca Reali and Emanuele Berti and Alessandra Corsi and B. S. Sathyaprakash and Digvijay Wadekar},
      year={2026},
      eprint={2604.11871},
      archivePrefix={arXiv},
      primaryClass={gr-qc},
      url={https://arxiv.org/abs/2604.11871}, 
}

@article{10.1093/mnras/staa201,
    author = {Aksaker, N and Yerli, S K and Erdoğan, M A and Kurt, Z and Kaba, K and Bayazit, M and Yesilyaprak, C},
    title = {Global Site Selection for Astronomy},
    journal = {Monthly Notices of the Royal Astronomical Society},
    volume = {493},
    number = {1},
    pages = {1204-1216},
    year = {2020},
    month = {03},
    issn = {0035-8711},
    doi = {10.1093/mnras/staa201},
    url = {https://doi.org/10.1093/mnras/staa201},
    eprint = {https://academic.oup.com/mnras/article-pdf/493/1/1204/32634447/staa201.pdf},
}

@techreport{fcc2025vol3,author={Benedikt, M. and Zimmermann, F. and others},title={Future Circular Collider Feasibility Study Report: Volume 3,
Civil Engineering, Implementation and Sustainability},institution={CERN},reportnumber={CERN-FCC-ACC-2025-0003},year={2025},url={https://arxiv.org/abs/2505.00273}
}

@techreport{ITERJASS2003,
  author       = {{ITER Joint Assessment of Specific Sites (JASS) Ad Hoc Group}},
  title        = {{JASS Final Report: ITER Candidate Sites}},
  institution  = {{ITER}},
  year         = {2003},
  month        = jan,
  number       = {V2.2},
  date         = {2003-01-26},
  url          = {https://www.mofa.go.jp/policy/s_tech/iter/jass0302.pdf},
  note         = {Final report of the Joint Assessment of Specific Sites for the ITER candidate sites: Clarington, Canada; Rokkasho, Japan; Cadarache, France; and Vandell{\`o}s, Spain.}
}

@misc{AJenkinsCENight,
 author  = "Andrew Jenkins",
 title   = {{Cosmic Explorer---Corner Station Nighttime Visualization}},
 year    = "2025",
    note = {{Accessed: 04-24-26}},
 url     = "https://commons.wikimedia.org/wiki/File:Cosmic_Explorer_-_Corner_Station_Nighttime_Visualization.jpg",
}

@misc{EPSG:5070,
    title = {{EPSG:5070}},
    url = {https://epsg.io/5070},
    note = {{Accessed: 04-20-26}}
}

@misc{NLCD2021,
title={{National Land Cover Database (NLCD) 2021 Products: U.S. Geological Survey data release}},
author = {Dewitz, J},
year = {2023},
doi = {https://doi.org/10.5066/P9JZ7AO3},
}

@article{PhysRevD.108.123009,
  title = {Cryogenic payloads for the Einstein Telescope: Baseline design with heat extraction, suspension thermal noise modeling, and sensitivity analyses},
  author = {Koroveshi, Xhesika and Busch, Lennard and Majorana, Ettore and Puppo, Paola and Rapagnani, Piero and Ricci, Fulvio and Ruggi, Paolo and Grohmann, Steffen},
  journal = {Phys. Rev. D},
  volume = {108},
  issue = {12},
  pages = {123009},
  numpages = {21},
  year = {2023},
  month = {Dec},
  publisher = {American Physical Society},
  doi = {10.1103/PhysRevD.108.123009},
  url = {https://link.aps.org/doi/10.1103/PhysRevD.108.123009}
}

@misc{MTDB,
title={{Master Address File/Topologically Integrated Geographic Encoding and Referencing
(MAF/TIGER) Database (MTDB)}},
author = {{U.S. Census Bureau}},
year = {2024},
}

@ARTICLE{2021arXiv210909882E,
       author = {{Evans}, Matthew and {Adhikari}, Rana X and {Afle}, Chaitanya and {Ballmer}, Stefan W. and {Biscoveanu}, Sylvia and {Borhanian}, Ssohrab and {Brown}, Duncan A. and {Chen}, Yanbei and {Eisenstein}, Robert and {Gruson}, Alexandra and {Gupta}, Anuradha and {Hall}, Evan D. and {Huxford}, Rachael and {Kamai}, Brittany and {Kashyap}, Rahul and {Kissel}, Jeff S. and {Kuns}, Kevin and {Landry}, Philippe and {Lenon}, Amber and {Lovelace}, Geoffrey and {McCuller}, Lee and {Ng}, Ken K.~Y. and {Nitz}, Alexander H. and {Read}, Jocelyn and {Sathyaprakash}, B.~S. and {Shoemaker}, David H. and {Slagmolen}, Bram J.~J. and {Smith}, Joshua R. and {Srivastava}, Varun and {Sun}, Ling and {Vitale}, Salvatore and {Weiss}, Rainer},
        title = "{A Horizon Study for Cosmic Explorer: Science, Observatories, and Community}",
      journal = {arXiv e-prints},
         year = 2021,
        month = sep,
          eid = {arXiv:2109.09882},
        pages = {arXiv:2109.09882},
          doi = {10.48550/arXiv.2109.09882},
archivePrefix = {arXiv},
       eprint = {2109.09882},
 primaryClass = {astro-ph.IM},
       adsurl = {https://ui.adsabs.harvard.edu/abs/2021arXiv210909882E}
}

@article{Datrier:2025wjs,
doi = {10.1088/1742-6596/3177/1/012093},
url = {https://doi.org/10.1088/1742-6596/3177/1/012093},
year = {2026},
month = {feb},
publisher = {IOP Publishing},
volume = {3177},
number = {1},
pages = {012093},
author = {Datrier, Laurence and Lovelace, Geoffrey and Smith, Joshua R. and Saenz, Andrew and Romero, Amber and the Cosmic Explorer Project},
title = {Site Evaluation and Cost Estimation for Cosmic Explorer},
journal = {Journal of Physics: Conference Series}
}

@article{avirgo,
doi = {10.1088/0264-9381/32/2/024001},
url = {https://doi.org/10.1088/0264-9381/32/2/024001},
year = {2014},
month = {dec},
publisher = {IOP Publishing},
volume = {32},
number = {2},
pages = {024001},
author = {Acernese, F and Agathos, M and Agatsuma, K and Aisa, D and Allemandou, N and Allocca, A and Amarni, J and Astone, P and Balestri, G and Ballardin, G and Barone, F and Baronick, J-P and Barsuglia, M and Basti, A and Basti, F and Bauer, Th S and Bavigadda, V and Bejger, M and Beker, M G and Belczynski, C and Bersanetti, D and Bertolini, A and Bitossi, M and Bizouard, M A and Bloemen, S and Blom, M and Boer, M and Bogaert, G and Bondi, D and Bondu, F and Bonelli, L and Bonnand, R and Boschi, V and Bosi, L and Bouedo, T and Bradaschia, C and Branchesi, M and Briant, T and Brillet, A and Brisson, V and Bulik, T and Bulten, H J and Buskulic, D and Buy, C and Cagnoli, G and Calloni, E and Campeggi, C and Canuel, B and Carbognani, F and Cavalier, F and Cavalieri, R and Cella, G and Cesarini, E and Mottin, E Chassande- and Chincarini, A and Chiummo, A and Chua, S and Cleva, F and Coccia, E and Cohadon, P-F and Colla, A and Colombini, M and Conte, A and Coulon, J-P and Cuoco, E and Dalmaz, A and D’Antonio, S and Dattilo, V and Davier, M and Day, R and Debreczeni, G and Degallaix, J and Deléglise, S and Pozzo, W Del and Dereli, H and Rosa, R De and Fiore, L Di and Lieto, A Di and Virgilio, A Di and Doets, M and Dolique, V and Drago, M and Ducrot, M and Endrőczi, G and Fafone, V and Farinon, S and Ferrante, I and Ferrini, F and Fidecaro, F and Fiori, I and Flaminio, R and Fournier, J-D and Franco, S and Frasca, S and Frasconi, F and Gammaitoni, L and Garufi, F and Gaspard, M and Gatto, A and Gemme, G and Gendre, B and Genin, E and Gennai, A and Ghosh, S and Giacobone, L and Giazotto, A and Gouaty, R and Granata, M and Greco, G and Groot, P and Guidi, G M and Harms, J and Heidmann, A and Heitmann, H and Hello, P and Hemming, G and Hennes, E and Hofman, D and Jaranowski, P and Jonker, R J G and Kasprzack, M and Kéfélian, F and Kowalska, I and Kraan, M and Królak, A and Kutynia, A and Lazzaro, C and Leonardi, M and Leroy, N and Letendre, N and Li, T G F and Lieunard, B and Lorenzini, M and Loriette, V and Losurdo, G and Magazzù, C and Majorana, E and Maksimovic, I and Malvezzi, V and Man, N and Mangano, V and Mantovani, M and Marchesoni, F and Marion, F and Marque, J and Martelli, F and Martellini, L and Masserot, A and Meacher, D and Meidam, J and Mezzani, F and Michel, C and Milano, L and Minenkov, Y and Moggi, A and Mohan, M and Montani, M and Morgado, N and Mours, B and Mul, F and Nagy, M F and Nardecchia, I and Naticchioni, L and Nelemans, G and Neri, I and Neri, M and Nocera, F and Pacaud, E and Palomba, C and Paoletti, F and Paoli, A and Pasqualetti, A and Passaquieti, R and Passuello, D and Perciballi, M and Petit, S and Pichot, M and Piergiovanni, F and Pillant, G and Piluso, A and Pinard, L and Poggiani, R and Prijatelj, M and Prodi, G A and Punturo, M and Puppo, P and Rabeling, D S and Rácz, I and Rapagnani, P and Razzano, M and Re, V and Regimbau, T and Ricci, F and Robinet, F and Rocchi, A and Rolland, L and Romano, R and Rosińska, D and Ruggi, P and Saracco, E and Sassolas, B and Schimmel, F and Sentenac, D and Sequino, V and Shah, S and Siellez, K and Straniero, N and Swinkels, B and Tacca, M and Tonelli, M and Travasso, F and Turconi, M and Vajente, G and van Bakel, N and van Beuzekom, M and van den Brand, J F J and Van Den Broeck, C and van der Sluys, M V and van Heijningen, J and Vasúth, M and Vedovato, G and Veitch, J and Verkindt, D and Vetrano, F and Viceré, A and Vinet, J-Y and Visser, G and Vocca, H and Ward, R and Was, M and Wei, L-W and Yvert, M and żny, A Zadro and Zendri, J-P},
title = {Advanced Virgo: a second-generation interferometric gravitational wave detector},
journal = {Classical and Quantum Gravity}
}

@article{LIGOScientific:2017vwq,
    author = "Abbott, B. P. and others",
    collaboration = "LIGO Scientific, Virgo",
    title = "{GW170817: Observation of Gravitational Waves from a Binary Neutron Star Inspiral}",
    eprint = "1710.05832",
    archivePrefix = "arXiv",
    primaryClass = "gr-qc",
    reportNumber = "LIGO-P170817",
    doi = "10.1103/PhysRevLett.119.161101",
    journal = "Phys. Rev. Lett.",
    volume = "119",
    number = "16",
    pages = "161101",
    year = "2017"
}

@techreport{ToobaSURF,
    author = {Tooba Ansar and Robert Schofield and Michael Landry},
    title = {A GIS-based workflow for the remote evaluation of potential Cosmic Explorer sites},
    institution = {LIGO},
    number = {T2500265},
    year = 2025,
    note = {Available at \url{https://dcc.ligo.org/LIGO-T2500265/public}}
}

@article{HASSAN201776,
title = {Monte Carlo performance studies for the site selection of the Cherenkov Telescope Array},
journal = {Astroparticle Physics},
volume = {93},
pages = {76-85},
year = {2017},
issn = {0927-6505},
doi = {https://doi.org/10.1016/j.astropartphys.2017.05.001},
url = {https://www.sciencedirect.com/science/article/pii/S0927650517300087},
author = {T. Hassan and L. Arrabito and K. Bernlöhr and J. Bregeon and J. Cortina and P. Cumani and F. {Di Pierro} and D. Falceta-Goncalves and R.G. Lang and J. Hinton and T. Jogler and G. Maier and A. Moralejo and A. Morselli and C.J. {Todero Peixoto} and M. Wood}
}

@article{Schoeck_2009,
doi = {10.1086/599287},
url = {https://doi.org/10.1086/599287},
year = {2009},
month = {may},
publisher = {University of Chicago Press},
volume = {121},
number = {878},
pages = {384},
author = {Schöck, M. and Els, S. and Riddle, R. and Skidmore, W. and Travouillon, T. and Blum, R. and Bustos, E. and Chanan, G. and Djorgovski, S. G. and Gillett, P. and Gregory, B. and Nelson, J. and Otárola, A. and Seguel, J. and Vasquez, J. and Walker, A. and Walker, D. and Wang, L.},
title = {Thirty Meter Telescope Site Testing I: Overview},
journal = {Publications of the Astronomical Society of the Pacific}
}

@article{LIGOScientific:2017ync,
    author = "Abbott, B. P. and others",
    collaboration = "LIGO Scientific, Virgo, Fermi GBM, INTEGRAL, IceCube, AstroSat Cadmium Zinc Telluride Imager Team, IPN, Insight-Hxmt, ANTARES, Swift, AGILE Team, 1M2H Team, Dark Energy Camera GW-EM, DES, DLT40, GRAWITA, Fermi-LAT, ATCA, ASKAP, Las Cumbres Observatory Group, OzGrav, DWF (Deeper Wider Faster Program), AST3, CAASTRO, VINROUGE, MASTER, J-GEM, GROWTH, JAGWAR, CaltechNRAO, TTU-NRAO, NuSTAR, Pan-STARRS, MAXI Team, TZAC Consortium, KU, Nordic Optical Telescope, ePESSTO, GROND, Texas Tech University, SALT Group, TOROS, BOOTES, MWA, CALET, IKI-GW Follow-up, H.E.S.S., LOFAR, LWA, HAWC, Pierre Auger, ALMA, Euro VLBI Team, Pi of Sky, Chandra Team at McGill University, DFN, ATLAS Telescopes, High Time Resolution Universe Survey, RIMAS, RATIR, SKA South Africa/MeerKAT",
    title = "{Multi-messenger Observations of a Binary Neutron Star Merger}",
    eprint = "1710.05833",
    archivePrefix = "arXiv",
    primaryClass = "astro-ph.HE",
    reportNumber = "LIGO-P1700294, VIR-0802A-17, FERMILAB-PUB-17-478-A-AE-CD",
    doi = "10.3847/2041-8213/aa91c9",
    journal = "Astrophys. J. Lett.",
    volume = "848",
    number = "2",
    pages = "L12",
    year = "2017"
}

@Inbook{Schilizzi2024,
author="Schilizzi, Richard T.
and Ekers, Ronald D.
and Dewdney, Peter E.
and Crosby, Philip",
title="Site Selection Story, 2006--2012: Decision",
bookTitle="The Square Kilometre Array: A Science Mega-Project in the Making, 1990-2012",
year="2024",
publisher="Springer International Publishing",
address="Cham",
pages="449--511",
isbn="978-3-031-51374-9",
doi="10.1007/978-3-031-51374-9_8",
url="https://doi.org/10.1007/978-3-031-51374-9_8"
}

@Article{app11188666,
AUTHOR = {Shang, Yan-Jun and Yang, Chang-Gen and Jin, Wei-Jun and Chen, Yan-Wei and Hasan, Muhammad and Wang, Yue and Li, Kun and Lin, Da-Ming and Zhou, Min},
TITLE = {Application of Integrated Geophysical Methods for Site Suitability of Research Infrastructures (RIs) in China},
JOURNAL = {Applied Sciences},
VOLUME = {11},
YEAR = {2021},
NUMBER = {18},
ARTICLE-NUMBER = {8666},
URL = {https://www.mdpi.com/2076-3417/11/18/8666},
ISSN = {2076-3417},
DOI = {10.3390/app11188666}
}

@article{Abreu:2023WI,
  author = "Abreu, Pedro  and  Albert, Andrea  and  Alfaro, Ruben Jose  and  Alfonso, Alexander  and  Alvarez, César  and  An, Qi  and  Angüner, Ekrem Oguzhan  and  Arcaro, Cornelia  and  Arceo, Roberto  and  Arias, Sandro  and  Arnaldi, Horacio  and  Assis, Pedro  and  Ayala Solares, Hugo Alberto  and  Bakalova, Alena  and  Barres de Almeida, Ulisses  and  Batković, Ivana  and  Bazo, Jose  and  Bellido, Jose  and  Belmont, Ernesto   and  BenZvi, Segev  and  Bernal, Abel  and  Bian, Wenyi  and  Bigongiari, Ciro  and  Bottacini, E.  and  Brogueira, Pedro  and  Bulik, Tomasz  and  Busetto, Giovanni  and  Caballero-Mora, Karen Salomé  and  Camarri, Paolo  and  Campos, Silvina  and  Cao, Wenyu  and  Cao, Zhe  and  Cao, Zhen  and  Capistrán, Tomás  and  Cardillo, Martina  and  Carquin, Edson  and  Carramiñana, Alberto  and  Castromonte, Cesar  and  Chang, Jinfan  and  Chaparro, Oscar  and  Chen, Shangming  and  Chianese, Marco  and  Chiavassa, Andrea  and  Chytka, Ladislav  and  Colallillo, Roberta  and  Conceição, Rúben  and  Consolati, Giovanni  and  Cordero, Raul  and  Costa, Pedro  and  Cotzomi, Jorge  and  Dasso, Sergio  and  De Angelis, Alessandro  and  Desiati, Paolo  and  Di Pierro, Federico  and  Di Sciascio, Giuseppe  and  Díaz Vélez, Juan Carlos  and  Dib, Claudio  and  Dingus, Brenda  and  Djuvsland, Julia Isabel  and  Dobrigkeit, Carola  and  Domingues Mendes, Luis Miguel  and  Dorigo, Tommaso  and  Doro, Michele  and  dos Reis, Alberto Corrêa  and  Du Vernois, Michael  and  Echiburu, Mauricio  and  Elsaesser, Dominik  and  Engel, Kristi  and  Ergin, Tulun  and  Espinoza, F.  and  Fang, Ke  and  Farfán Carreras, Fernando  and  Fazzi, Alberto  and  Feng, Cunfeng  and  Feroci, Marco  and  Fraija, Nissim  and  Fraija, Sara  and  Franceschini, Alberto  and  Franco, G. F.  and  Funk, Stefan  and  Garcia, Sayri  and  Garcia-Gonzalez, Jose Andres  and  Garfias, Fernando  and  Giacinti, Gwenael  and  Gibilisco, Lucio  and  Glombitza, Jonas  and  Goksu, Hazal  and  Gong, Guanghua  and  González, Borja Serrano  and  Gonzalez, Magda  and  Goodman, Jordan A.  and  Gu, Minhao  and  Guarino, Fausto  and  Gupta, Shivangi  and  Haist, Fabian  and  Hakobyan, hayk  and  Han, Guangchao  and  Hansen, Patricia María  and  Harding, J. Patrick  and  Helo, Juan  and  Herzog, Ian  and  d. Hidalgo, Hugo  and  Hinton, Jim  and  Hu, Kun  and  Huang, Dezhi  and  Huentemeyer, Petra  and  Hueyotl-Zahuantitla, Filiberto  and  Iriarte, Arturo  and  Isaković, Jasmina  and  Isolia, Antonio  and  Joshi, Vikas  and  Jurysek, Jakub  and  Kaci, Samy  and  Kieda, Dave  and  La Monaca, Fabio  and  La Mura, Giovanni  and  Lang, Rodrigo Guedes  and  Laspiur, Roxana  and  Lavitola, Luigi  and  LEE, Jason  and  Leitl, Franziska  and  Lessio, Luigi  and  Li, Cong  and  Li, Jian  and  Li, Kai  and  Li, Tianyang  and  Liberti, Barbara  and  Lin, Sujie  and  Liu, Dong  and  Liu, Jia  and  Liu, Ruoyu  and  Longo, Francesco  and  Luo, Yu  and  Lv, Jing  and  Macerata, Elena  and  Malone, Kelly  and  Mandat, Dusan  and  Manganaro, Marina  and  Mariani, Mario  and  Mariazzi, Analisa  and  Mariotti, Mosè  and  Marrodan, Teresa  and  Martínez, Jesús  and  Martínez-Huerta, Humberto -  and  Medina, S.  and  Melo, Diego  and  Mendes, Filipe  and  Meza, Erick  and  Miceli, Davide  and  Miozzi, Silvia  and  Mitchell, Alison  and  Molinario, Andrea  and  Morales-Olivares, Oscar G.  and  Moreno, Eduardo  and  Morselli, Aldo  and  Mossini, Eros  and  Mostafa, Miguel  and  Muleri, Fabio  and  Nardi, F.  and  Negro, A.  and  Nellen, Lukas  and  Novotný, Vladimír  and  Orlando, Elena  and  Osorio, Mabel  and  Otiniano, Luis  and  Peresano, Michele  and  Piano, Giovanni  and  Pichel, Ana  and  Pihet, Marine  and  Pimenta, Mário  and  Prandini, Elisa  and  Qin, Jiajun  and  Quispe, Erick  and  Raino, Silvia  and  Rangel, E.  and  Reisenegger, Andreas  and  Ren, Helena X.  and  Rescic, Filip  and  Reville, Brian  and  Rho, Chang Dong  and  Riquelme, Mario  and  Rodriguez Fernandez, Gonzalo  and  Roh, Youn  and  Romero, Gustavo E.  and  ROSSI, Biagio  and  Rovero, Adrian C.  and  Ruiz-Velasco, Edna  and  Salazar, Germán  and  Samanes, Jorge  and  Sánchez, Federico Andrés  and  Sandoval, Andrés  and  Santander, Marcos  and  Santonico, Rinaldo  and  Santos, Gizele Lian P.  and  Saviano, Ninetta  and  Schneider, Michael  and  Schneider, Martin  and  Schoorlemmer, Harm  and  Serna-Franco, José Erandi  and  Serrano, Victor  and  Smith, Andrew  and  Son, Youngwan  and  Soto, Orlando  and  Springer, Wayne Robert  and  Stuani, Luiz Augusto  and  Sun, Hao  and  Tang, Ruiyi  and  Tang, Zebo  and  Tapia, Sebastian  and  Tavani, Marco  and  Terzić, Tomislav  and  Tollefson, Kirsten  and  Tomé, Bernardo  and  Torres, Ibrahim Daniel  and  Torres-Escobedo, Ramiro  and  Trinchero, Gian Carlo  and  Turner, Rhiannon M.  and  Ulloa, Pablo  and  Valore, Laura  and  van Eldik, Christopher  and  Vergara, Indira  and  Viana, Aion  and  Vicha, Jakub  and  Vigorito, Carlo Francesco  and  Vittorini, Valerio  and  Wang, Bo  and  Wang, Jieshuang  and  Wang, Lingyu  and  Wang, Xiaojie  and  Wang, Xiangyu  and  Wang, Xinsheng  and  Wang, Zhen  and  Waqas, Muhammad  and  Watson, Ian James  and  Werner, Felix  and  White, Richard  and  Wiebusch, Christopher  and  Willox, Elijah J.  and  Wohlleben, Frederik  and  Wu, Sha  and  Xi, Shaoqiang  and  Xiao, Gang  and  Yang, Lili  and  Yang, Ruizhi  and  Yanyachi, Raul  and  Yao, Zhiguo  and  Zavrtanik, Danilo  and  Zhang, Hongfei  and  Zhang, Haiming  and  Zhang, Shaoru  and  Zhang, Xiaopeng  and  Zhang, Yuanping  and  Zhao, Jing  and  Zhao, Lei  and  Zhou, Zhou  and  Zhu, Chengguang  and  Zhu, Pengzhe  and  Zuo, Xiong",
  title = "{An update on site search activities for SWGO}",
  doi = "10.22323/1.444.0752",
  journal = "PoS",
  year = 2023,
  volume = "ICRC2023",
  pages = "752"
}

@article{aligo,
    author = "Aasi, J. and others",
    collaboration = "LIGO Scientific",
    title = "{Advanced LIGO}",
    eprint = "1411.4547",
    archivePrefix = "arXiv",
    primaryClass = "gr-qc",
    doi = "10.1088/0264-9381/32/7/074001",
    journal = "Class. Quant. Grav.",
    volume = "32",
    pages = "074001",
    year = "2015"
}

@article{LIGOScientific:2016aoc,
    author = "Abbott, B. P. and others",
    collaboration = "LIGO Scientific, Virgo",
    title = "{Observation of Gravitational Waves from a Binary Black Hole Merger}",
    eprint = "1602.03837",
    archivePrefix = "arXiv",
    primaryClass = "gr-qc",
    reportNumber = "LIGO-P150914",
    doi = "10.1103/PhysRevLett.116.061102",
    journal = "Phys. Rev. Lett.",
    volume = "116",
    number = "6",
    pages = "061102",
    year = "2016"
}

@article{
doi:10.1073/pnas.2012865117,
author = {Christoph Nolte },
title = {High-resolution land value maps reveal underestimation of conservation costs in the United States},
journal = {Proceedings of the National Academy of Sciences},
volume = {117},
number = {47},
pages = {29577-29583},
year = {2020},
doi = {10.1073/pnas.2012865117},
eprint = {https://www.pnas.org/doi/pdf/10.1073/pnas.2012865117}}

@misc{NationalMap,
title={{3D Elevation Program 1-Arc second Resolution Digital Elevation Model}},
author = {{U.S. Geological Survey}},
year = {2013-2024},
}

@misc{Kuns_2020,
    title={{Cosmic Explorer Site Search}},
  author = {Kuns, K. and Evans, M.},
  collaboration = {Cosmic Explorer},
  year = {2020},
  url = {https://dcc.cosmicexplorer.org/CE-T2000016/public}
}

@misc{celsTilt,
    title={{An Explanation of Arm Tilt for the Cosmic Explorer Location Search (CELS) Code}},
  author = {Smith, J.R. and Datrier, L. and Lovelace, G.},
  collaboration = {Cosmic Explorer},
  year = {2026},
  url = {https://dcc.cosmicexplorer.org/CE-T2600008}
}

@misc{CELS_codebase,
    title = {{Cosmic Explorer Location Search code}},
    author = {Lovelace, Geoffrey and Datrier, Laurence and Smith, Joshua},
    url = {https://gitlab.com/cosmic-explorer/cels},
    year = {2025}
}

@misc{CABridgeCosts,
    title = {{Comparative Bridge Costs}},
    author = {{State of California Department of Transportation}},
    url = {https://dot.ca.gov/-/media/dot-media/programs/local-assistance/documents/hbp/2024/comp-br-costs-2022.pdf},
    year = {2023}
}

@misc{FLDOTcosts,
    title = {{FDOT District 3 Estimated Transportation Costs }},
    author = {{State of Florida Department of Transportation}},
    url = {https://crtpa.org/files/120272339.pdf},
    year = {2026}
}

@misc{BLSInflationCalculator,
    title = {{CPI Inflation Calculator}},
    author = {{US Bureau of Labor Statistics}},
    url = {https://www.bls.gov/data/inflation_calculator.htm},
    year = {2026}
}

@article{10.1063/5.0242016,
    author = {Daniel, Kathryne J. and Smith, Joshua R. and Ballmer, Stefan and Bristol, Warren and Driggers, Jennifer C. and Effler, Anamaria and Evans, Matthew and Hoover, Joseph and Kuns, Kevin and Landry, Michael and Lovelace, Geoffrey and Lukinbeal, Chris and Mandic, Vuk and Pham, Kiet and Read, Jocelyn and Russell, Joshua B. and Schiettekatte, François and Schofield, Robert M. S. and Scholz, Christopher A. and Shoemaker, David H. and Sledge, Piper and Strunk, Amber},
    title = {{Criteria for identifying and evaluating locations that could potentially host the Cosmic Explorer observatories}},
    journal = {Review of Scientific Instruments},
    volume = {96},
    number = {1},
    pages = {014502},
    year = {2025},
    month = {01},
    issn = {0034-6748},
    doi = {10.1063/5.0242016},
    url = {https://doi.org/10.1063/5.0242016},
    eprint = {https://pubs.aip.org/aip/rsi/article-pdf/doi/10.1063/5.0242016/20339208/014502\_1\_5.0242016.pdf},
}

@article{Amann:2020jgo,
    author = "Amann, Florian and others",
    title = "{Site-selection criteria for the Einstein Telescope}",
    eprint = "2003.03434",
    archivePrefix = "arXiv",
    primaryClass = "physics.ins-det",
    doi = "10.1063/5.0018414",
    journal = "Rev. Sci. Instrum.",
    volume = "91",
    number = "9",
    pages = "9",
    year = "2020"
}

@misc{KalogeraReport,
    title = {{NSF MPS AC Subcommittee on Next-Generation Gravitational-Wave Detector Concepts Report, March 2024}},
    author = {Kalogera, Vicky and Alexander, Stephon and Aprahamian, Ani and Bennett, Charles L and Branchesi, Marica and Cominsky, Lynn R and  Halkiadakis, Eva and Gonzalez, Gabriela and L\"{u}ck, Harald and Scholberg, Kate and Wechsler, Risa and Will, Clifford M and Zaldarriaga, Matias},
    year = {2024},
    url = {https://nsf-gov-resources.nsf.gov/files/mpsac-nggw-subcommittee-repor-2024-03-23-r.pdf?VersionId=uhdtblJUPYsavII5EUuSNQJcFdRr92q_}
}

@article{LIGOScientific:2007fwp,
    author = "Abbott, B. P. and others",
    collaboration = "LIGO Scientific",
    title = "{LIGO: The Laser interferometer gravitational-wave observatory}",
    eprint = "0711.3041",
    archivePrefix = "arXiv",
    primaryClass = "gr-qc",
    reportNumber = "LIGO-P070082-04",
    doi = "10.1088/0034-4885/72/7/076901",
    journal = "Rept. Prog. Phys.",
    volume = "72",
    pages = "076901",
    year = "2009"
}

@article{Punturo:2010zz,
    author = "Punturo, M. and others",
    editor = "Ricci, Fulvio",
    title = "{The Einstein Telescope: A third-generation gravitational wave observatory}",
    doi = "10.1088/0264-9381/27/19/194002",
    journal = "Class. Quant. Grav.",
    volume = "27",
    pages = "194002",
    year = "2010"
}

@misc{evans2023cosmicexplorersubmissionnsf,
      title={{Cosmic Explorer: A Submission to the NSF MPSAC ngGW Subcommittee}}, 
      author={Matthew Evans and Alessandra Corsi and Chaitanya Afle and Alena Ananyeva and K. G. Arun and Stefan Ballmer and Ananya Bandopadhyay and Lisa Barsotti and Masha Baryakhtar and Edo Berger and Emanuele Berti and Sylvia Biscoveanu and Ssohrab Borhanian and Floor Broekgaarden and Duncan A. Brown and Craig Cahillane and Lorna Campbell and Hsin-Yu Chen and Kathryne J. Daniel and Arnab Dhani and Jennifer C. Driggers and Anamaria Effler and Robert Eisenstein and Stephen Fairhurst and Jon Feicht and Peter Fritschel and Paul Fulda and Ish Gupta and Evan D. Hall and Giles Hammond and Otto A. Hannuksela and Hannah Hansen and Carl-Johan Haster and Keisi Kacanja and Brittany Kamai and Rahul Kashyap and Joey Shapiro Key and Sanika Khadkikar and Antonios Kontos and Kevin Kuns and Michael Landry and Philippe Landry and Brian Lantz and Tjonnie G. F. Li and Geoffrey Lovelace and Vuk Mandic and Georgia L. Mansell and Denys Martynov and Lee McCuller and Andrew L. Miller and Alexander Harvey Nitz and Benjamin J. Owen and Cristiano Palomba and Jocelyn Read and Hemantakumar Phurailatpam and Sanjay Reddy and Jonathan Richardson and Jameson Rollins and Joseph D. Romano and Bangalore S. Sathyaprakash and Robert Schofield and David H. Shoemaker and Daniel Sigg and Divya Singh and Bram Slagmolen and Piper Sledge and Joshua Smith and Marcelle Soares-Santos and Amber Strunk and Ling Sun and David Tanner and Lieke A. C. van Son and Salvatore Vitale and Benno Willke and Hiro Yamamoto and Michael Zucker},
      year={2023},
      eprint={2306.13745},
      archivePrefix={arXiv},
      primaryClass={astro-ph.IM},
      url={https://arxiv.org/abs/2306.13745}, 
}

@article{Anthony:2022,
    author = {Anthony, Robert E. and Ringler, Adam T. and Wilson, David C.},
    title = "{Seismic Background Noise Levels across the Continental United States from USArray Transportable Array: The Influence of Geology and Geography}",
    journal = {Bulletin of the Seismological Society of America},
    volume = {112},
    number = {2},
    pages = {646-668},
    year = {2022},
    month = {01},
    issn = {0037-1106},
    doi = {10.1785/0120210176},
    url = {https://doi.org/10.1785/0120210176},
    eprint = {https://pubs.geoscienceworld.org/ssa/bssa/article-pdf/112/2/646/5638811/bssa-2021176.1.pdf},
}

@article{Saleem_2022,
doi = {10.1088/1361-6382/ac3b99},
url = {https://doi.org/10.1088/1361-6382/ac3b99},
year = {2021},
month = {dec},
publisher = {IOP Publishing},
volume = {39},
number = {2},
pages = {025004},
author = {Saleem, M and Rana, Javed and Gayathri, V and Vijaykumar, Aditya and Goyal, Srashti and Sachdev, Surabhi and Suresh, Jishnu and Sudhagar, S and Mukherjee, Arunava and Gaur, Gurudatt and Sathyaprakash, Bangalore and Pai, Archana and Adhikari, Rana X and Ajith, P and Bose, Sukanta},
title = {The science case for LIGO-India},
journal = {Classical and Quantum Gravity}
}

@unpublished{Joey,
  author = {Key, J S and , Han, R J and Ara, S R},
  title = {{Cosmic Explorer Observatory Conceptual Design}},
  institution = {University of Washington},
  note = {{J. Phys.: Conf. Series, submitted; \url{https://dcc.cosmicexplorer.org/CE-P2500001/public}}},
  year = {2025}
}

@article{Bristol_2026,
doi = {10.1088/1742-6596/3177/1/012092},
url = {https://doi.org/10.1088/1742-6596/3177/1/012092},
year = {2026},
month = {feb},
publisher = {IOP Publishing},
volume = {3177},
number = {1},
pages = {012092},
author = {Bristol, Warren and Lukinbeal, Chris and Hoover, Joseph and Daniel, Kate and Sledge, Piper and Russell, Joshua and Evans, Matthew and on behalf of the Cosmic Explorer Project},
title = {An Interdisciplinary Site Suitability Analysis for Cosmic Explorer},
journal = {Journal of Physics: Conference Series}
}

@article{Shoemaker_2026,
doi = {10.1088/1742-6596/3177/1/012091},
url = {https://doi.org/10.1088/1742-6596/3177/1/012091},
year = {2026},
month = {feb},
publisher = {IOP Publishing},
volume = {3177},
number = {1},
pages = {012091},
author = {Shoemaker, David H. and on behalf of the Cosmic Explorer Project},
title = {Cosmic Explorer Project Status},
journal = {Journal of Physics: Conference Series}
}

\end{document}